\documentclass[prd, twocolumn,letterpaper, amsmath, amssymb, preprintnumbers, showpacs, floatfix, nofootinbib]{revtex4-1}
\usepackage[utf8]{inputenc}

\usepackage{graphicx}
\usepackage{subfigure}
\usepackage{float}
\usepackage{color}
\usepackage{comment}
\usepackage[dvipsnames]{xcolor} \pdfoutput=1
\allowdisplaybreaks 
\usepackage[hypertexnames=false]{hyperref}
\hypersetup{
    colorlinks=true, 
    linkcolor=blue,  
    citecolor=blue,     
    filecolor=blue,   
    urlcolor=blue       
}
\usepackage{slashed}
\usepackage{mathtools}

\newcommand{\hethree}{{}^3\text{He}}
\newcommand{\hthree}{{}^3\text{H}}

\newcommand{\ket}[1]{\left| #1 \right>} 

\newcommand{\pislash}[0]{\pi\!\!\!/}

\begin{document}

\title{Few-nucleon matrix elements in pionless effective field theory in a finite volume}
\author{W. Detmold}
\author{P. E. Shanahan}
\affiliation{
	Center for Theoretical Physics, 
	Massachusetts Institute of Technology, 
	Cambridge, MA 02139, USA}
\affiliation{The NSF AI Institute for Artificial Intelligence and Fundamental Interactions}
 
\date{\today}

\begin{abstract}
\preprint{MIT-CTP/5275}
    Pionless effective field theory in a finite volume (FVEFT$_{\pislash}$) is investigated as a framework for the analysis of multi-nucleon spectra and matrix elements calculated in lattice QCD (LQCD). By combining FVEFT$_{\pislash}$ with the stochastic variational method, the spectra of nuclei with atomic number $A\in\{2,3\}$ are matched to existing finite-volume LQCD calculations at heavier-than-physical quark masses corresponding to a pion mass $m_\pi=806$ MeV, thereby enabling infinite-volume binding energies to be determined using infinite-volume variational calculations. Based on the variational wavefunctions that are constructed in this approach, the finite-volume matrix elements of various local operators are computed in FVEFT$_{\pislash}$ and matched to LQCD calculations of the corresponding QCD operators in the same volume, thereby determining the relevant one and two-body EFT counterterms and enabling an extrapolation of the LQCD matrix elements to infinite volume. As examples, the scalar, tensor, and axial matrix elements are considered, as well as the magnetic moments and the isovector longitudinal momentum fraction. 
\end{abstract}

\maketitle

\section{Introduction}

Over the last 30 years, effective field theories have revolutionised nuclear physics, systematising the study of nucleon-nucleon interactions and the properties of light nuclei.
Pionless effective field theory for few-nucleon systems (EFT$_{\pislash}$) in particular, which focuses on momenta below the pion mass, has emerged as a powerful tool with which to understand low-energy nuclear processes in many contexts~\cite{Kaplan:1998tg,Kaplan:1998we,vanKolck:1998bw,Chen:1999tn,Bedaque:2002mn} (see Ref.~\cite{Hammer:2019poc} for a recent review). 
Notably, in addition to its use in the analysis of experimental data, EFT$_{\pislash}$ has found a key role in the analysis of  lattice Quantum Chromodynamics (LQCD) calculations of nuclear systems, providing a direct bridge between QCD and nuclear physics.

For example, in Ref.~\cite{Barnea:2013uqa}, LQCD calculations of $A\in\{2,3\}$ nuclei at heavier-than-physical quark masses~\cite{Beane:2012ey} were matched to auxiliary field diffusion Monte-Carlo calculations with EFT$_{\pislash}$ interactions to constrain the two and three-body counterterms of the EFT; these were then used to make predictions for larger nuclei of atomic number $A\leq 6$.
Further developments were presented in Ref.~\cite{Kirscher:2015yda}, and this approach was extended to the next order in the EFT, and to still larger nuclei, in Refs.~\cite{Contessi:2017rww} and \cite{Bansal:2017pwn}. 
Ref.~\cite{Kirscher:2017fqc} presents studies of the quark-mass dependence of the magnetic moments and polarisabilities of $A\in\{2,3\}$ nuclear systems, with both experimental results and LQCD calculations at larger-than-physical values of the pion mass used to constrain the counterterms of EFT$_{\pislash}$.
Similarly, an early application of finite-volume EFT$_{\pislash}$ to electroweak matrix elements was presented in Ref.~\cite{Detmold:2004qn} and extended in Ref.~\cite{Briceno:2012yi}.
Furthermore, EFT$_{\pislash}$ provided a powerful approach to analysing LQCD calculations of second order weak processes \cite{Shanahan:2017bgi,Tiburzi:2017iux,Davoudi:2020gxs,Davoudi:2020xdv,Briceno:2019opb}.

Recently,  Eliyahu {\it et al.}~\cite{Eliyahu:2019nkz} have taken the next steps in  this approach and used EFT$_{\pislash}$ implemented via the stochastic variational method (SVM)~\cite{Varga:1995dm,Mitroy:2013eom} in a finite cubic volume to analyse the binding energies of atomic number $A\in\{2,3\}$ systems in that finite volume. 
Since the effects of a finite volume manifest in the infrared domain, they can be captured in low-energy effective descriptions of QCD such as that provided by EFT$_{\pislash}$. Since LQCD calculations are typically performed in multiple volumes to enable an infinite-volume extrapolation, performing the EFT in the same volumes also maximises the constraining power of the LQCD results.  As input, Ref.~\cite{Eliyahu:2019nkz} utilised the NPLQCD collaboration's LQCD computations of the ground-state energies and finite-volume energy shifts of these systems in 3 different lattice volumes with spatial extents $L\in\{3.4,\,4.5,\,6.7\}$~fm at unphysical quark masses corresponding to the SU(3)$_f$ flavour-symmetric point where the up, down, and strange quark masses are degenerate and correspond to a  pion mass $m_\pi=m_K=806$ MeV~\cite{Beane:2012vq}. Using EFT$_{\pislash}$ in the same volumes to determine counterterms, the  binding energies were extrapolated to infinite volume.
As this exemplifies, the finite-volume pionless effective field theory (FVEFT$_{\pislash}$) approach provides a powerful alternative to L\"uscher's method~\cite{Luscher:1986pf}, in which finite-volume energies from LQCD calculations are used to determine infinite-volume scattering phase shifts and bound-state energies. While L\"uscher's method and its generalisations are model independent, the existing formalism is limited to two and three-particle systems. The matching of LQCD results to EFT$_{\pislash}$, on the other hand, requires an underlying EFT but can be applied to any system that the EFT can address.

In this work, the application of FVEFT$_{\pislash}$ is extended to nuclear matrix elements for the first time.
After a brief summary of the relevant EFT$_{\pislash}$ Lagrangians in Sec.~\ref{sec:pionless}, the SVM is introduced in Sec.~\ref{sec:svm}. Section~\ref{sec:groundstatematching} presents the results of tuning the relevant two and three-body counterterms of FVEFT to reproduce the ground-state energies of $A\in\{2,3\}$ nuclei computed in LQCD, paralleling the analysis of Ref.~\cite{Eliyahu:2019nkz}. Having determined these counterterms, Sec.~\ref{sec:ME} presents the tuning of the counterterms describing the interactions of external currents to reproduce LQCD calculations of nuclear matrix elements, thereby enabling an extrapolation of the finite-volume matrix elements to infinite volume. In particular, LQCD calculations of scalar, tensor, and axial matrix elements, as well as  magnetic moments and isovector longitudinal momentum fractions, of $A\in\{2,3\}$ states are investigated. 
To conclude, the outlook for these calculations is discussed in Sec.~\ref{sec:discuss}.

\section{Pionless effective field theory}
\label{sec:pionless}
The pionless EFT Lagrangian describing the low-energy interactions of nucleons is given by
\begin{equation}\label{eq:pionlessL}
\mathcal{L}=\mathcal{L}_{1}+\mathcal{L}_{2}+\mathcal{L}_{3}+\ldots,
\end{equation}
where 
\begin{equation}
\mathcal{L}_{1}=N^{\dagger}\left(i D_{0}+\frac{\mathbf{D}^{2}}{2 M_{N}}\right) N +\ldots
\label{eq:Lagrangian1body}
\end{equation}
contains the single-nucleon kinetic operator expanded in the non-relativistic (NR) limit. Here, $N$ represents the nucleon field, $M_N$ is the nucleon mass, and the ellipsis denotes higher-order terms. The leading-order two-nucleon interactions enter as 
\begin{align}
\mathcal{L}_{2}=&-{C_S} \left(N^{T} P_{i} N\right)^{\dagger}\left(N^{T} P_{i} N\right)
\nonumber \\
&
-{C_T}\left(N^{T} \overline{P}_{a} N\right)^{\dagger}\left(N^{T} \overline{P}_{a} N\right) +\ldots
\label{eq:Lagrangian2body}
\end{align}
where 
\begin{equation}
P_{i} \equiv \frac{1}{\sqrt{8}} \sigma_{2} \sigma_{i} \tau_{2}, \hspace{7mm}
\overline{P}_{a} \equiv \frac{1}{\sqrt{8}} \sigma_{2} \tau_{2} \tau_{a}
\label{eq:projector}
\end{equation}
are projectors onto spin-triplet and sin-singlet two-nucleon states respectively, and $C_S$ and $C_T$ are the relevant two-body  low-energy constants (LECs). Here  $\sigma_k$ and $\tau_a$ are the Pauli  matrices acting in spin and isospin space, respectively. Eq.~\eqref{eq:Lagrangian2body} can be re-expressed in a different basis as \cite{Mehen:1999qs} 
\begin{equation}
\mathcal{L}_{2}=-\frac{1}{2}\left[C_{0}\left(N^{\dagger} N\right)^{2}+C_{1}\left(N^{\dagger} \vec{\sigma} N\right)^{2} \right]  + \ldots,
\label{eq:Lagrangian2bodyalt}
\end{equation}
where
\begin{equation}
\label{eq:C01CSTrelation}
C_T=C_{0}-3 C_{1} \hspace{3mm}\text{and}\hspace{3mm} C_S=C_{0}+C_{1}.
\end{equation}
Three-body interactions naively enter at higher order, but must be promoted to leading order as argued in Refs.~\cite{Bedaque:1998kg,Bedaque:1998km}, and the relevant contribution to the Lagrangian is
\begin{equation}
    \mathcal{L}_{3} = - \frac{D_0}{6}(N^\dagger N)^3  + \ldots,
    \label{eq:Lagrangian3body}
\end{equation}
where $D_0$ is the leading-order three-body LEC.

\subsection{Weak interactions}

The weak decays and interactions of nuclear states arise, after integrating out the weak gauge boson, through the effective Lagrangian (valid for energies $E\ll M_W$)
\begin{equation}
\mathcal{L}_{W}=-\frac{G_{F}}{\sqrt{2}} l_{+}^{\mu} J_{\mu}^{-}+ \text{h.c.} +\cdots,
\end{equation}
where $G_F$ is the Fermi constant, $l_+^\mu$ involves a charged lepton and neutrino, and the hadronic weak current can be expressed in terms of the vector ($V_\mu$) and axial-vector ($A_\mu$) currents as
\begin{equation}
J_{i,\mu}=V_{i,\mu}-A_{i,\mu},
\end{equation}
with $J^\pm_\mu=J_{1,\mu}\pm i\ J_{2,\mu}$.
The isovector axial-vector current in EFT$_{\pislash}$ is given by~\cite{Butler:2001jj}\footnote{Note that the normalisation of $L_{1,A}$ used here is the same as in Ref.~\cite{Butler:2001jj} but differs from Ref.~\cite{Butler:1999sv}, which uses a different projector definition.}
\begin{align}
A_{i,a}=& \frac{g_{A}}{2} N^{\dagger} \tau_{a} \sigma_{i} N \nonumber \\
&+L_{1, A} \left(N^{T} P_{i} N\right)^{\dagger}\left(N^{T} \overline{P}_{a} N\right) + \text{h.c.}
+\ldots,
\label{eq:axial_current}
\end{align}
where the ellipsis denotes higher-order terms. In this expression, $g_A$ is the nucleon axial charge, and the term proportional to the two-nucleon LEC $L_{1,A}$ provides the next-to-leading-order (NLO) corrections to the $pp\to d e^+ \nu$ fusion process, for example.

The isoscalar axial current is similarly given by~\cite{Butler:1999sv}\footnote{Note that Ref.~\cite{Butler:1999sv} uses a different projector definition than here, and correspondingly a different normalisation of $L_{2,A}$.}
\begin{align}
A_{i,0}=& - \frac{g_{A,0}}{2} N^{\dagger}  \sigma_{i} N \nonumber \\
&- 2 i L_{2, A}\epsilon_{ijk}\left(N^{T} P_{j} N\right)^{\dagger}\left(N^{T} P_k N\right) +\cdots,
\label{eq:isoscalar_axial_current}
\end{align}
where the isoscalar nucleon axial charge is $g_{A,0}$, and the LEC  $L_{2, A}$ enters at the same order.

\subsection{Electromagnetic interactions}

In the presence of an external electromagnetic (EM) field, the Lagrangians defined above are modified such that the derivatives are replaced by EM-covariant derivatives $D_\mu=\partial_\mu+i Q A_\mu$, where $A_\mu$ is the vector potential and $Q$ is the electric charge operator, and also by the addition of terms depending on the magnetic field $\bf{B}$:\footnote{The notation of Ref.~\cite{Rupak:1999rk} is used.}
\begin{equation}
    {\cal L}_{1,\text{EM}} + {\cal L}_{2,\text{EM}} = J^{EM}_{i} {\bf B}_i, 
    \label{eq:Lmagfield}
\end{equation}
where the isoscalar and isovector currents coupling to the magnetic field are 
\begin{align}
    J^{EM}_{i}=&  \frac{e}{2 M_N} N^{\dagger}(\kappa_0 + \tau_3 \kappa_1) {\sigma}_i N \nonumber \\
    &{}-e L_{2} i \epsilon_{i j k }\left(N^{T} P_{k} N\right)^{\dagger}\left(N^{T} P_{j} N\right) \nonumber \\
       &{} +  e L_{1}\left(N^{T} P_{i} N\right)^{\dagger}\left(N^{T} \bar{P}_{3} N\right) +\text{h.c.},\label{eq:JEM}
\end{align}
where $L_1$ and $L_2$ are two-body LECs and
\begin{equation}
\kappa_{0}=\frac{1}{2}\left(\kappa_{p}+\kappa_{n}\right) \text { and } \kappa_{1}=\frac{1}{2}\left(\kappa_{p}-\kappa_{n}\right)
\end{equation}
are the isoscalar and isovector nucleon magnetic moments. Note that electric field contributions and EM three-body interactions enter at higher order. 

\subsection{Scalar and tensor currents}
\label{subsec:scalarEFT}

The isovector and isoscalar scalar currents that arise from Higgs couplings and from potential dark matter interactions are given by
\begin{align}
S_0 ={}& g_{S,0} N^{\dagger}  N 
-\widetilde{C}_S \left(N^{T} P_{i} N\right)^{\dagger}\left(N^{T} P_{i} N\right),
\nonumber \\
&
-\widetilde{C}_T\left(N^{T} \overline{P}_{a} N\right)^{\dagger}\left(N^{T} \overline{P}_{a} N\right) +\ldots,
\label{eq:scalar_isoscalar_current}
\\[10pt]
S_a ={}& g_{S,3} N^{\dagger} \tau_a  N 
\nonumber\\
&+i  \widetilde{C}_V \epsilon_{abc}\left(N^{T} \overline{P}_b  N\right)^{\dagger}\left(N^{T} \overline{P}_c N\right)
 +\ldots .
\label{eq:scalar_isovector_current}
\end{align}
Here $g_{S,0}$ and $g_{S,3}$ are the isoscalar and isovector one-body LECs that are related to the nucleon $\sigma$ terms.
As discussed in Ref.~\cite{Krebs:2020plh}, the two-body terms in the isoscalar scalar current are related to the corresponding terms in the strong Lagrangian, Eq.~\eqref{eq:Lagrangian2body}. In particular, the LECs $\widetilde{C}_{S,T}$ are the quark-mass--independent pieces of the Lagrangian couplings $C_{S,T}$. 

For the isoscalar and isovector antisymmetric tensor currents, the relevant EFT$_{\pislash}$ expressions are
\begin{align}
T_{ij,0} ={}& \frac{g_{T,0}}{2} \epsilon_{ijk} N^{\dagger} \sigma_{k} N  \nonumber\\
&+i L_{2, T}  \left(N^{T} P_{i} N\right)^{\dagger}\left(N^{T} P_j N\right) +\text{h.c.}+ \ldots,
\label{eq:tensor_isoscalar_current}
\\
T_{ij,a} ={}& \frac{g_{T,3}}{2} \epsilon_{ijk} N^{\dagger} \tau_{a} \sigma_{k} N \nonumber \\
&+ L_{1, T} \epsilon_{ijk} \left(N^{T} {P}_k  N\right)^{\dagger}\left(N^{T} \overline{P}_a N\right) + \text{h.c.}
 +\ldots,
\label{eq:tensor_isovector_current}
\end{align}
where the one and two-body isoscalar (isovector) tensor LECs are $g_{T,0}$ and $L_{2,T}$ ($g_{T,3}$ and $L_{1,T}$).

\subsection{Twist-two operators}

The unpolarised twist-two operators that define moments of parton distributions enter in EFT$_{\pislash}$ as \cite{Chen:2004zx,Beane:2004xf}
\begin{align}
O^{\mu_{0} \ldots \mu_{n}}=&\left\langle x^{n}\right\rangle_{0} v^{\mu_{0}} \cdots v^{\mu_{n}} N^{\dagger} N\left[1+\alpha_{n,0} N^{\dagger} N\right], \\ 
O^{\mu_{0} \ldots \mu_{n}}=&\left\langle x^{n}\right\rangle_{3} v^{\mu_{0}} \cdots v^{\mu_{n}} N^{\dagger}\tau_3 N\left[1+\alpha_{n,3} N^{\dagger} N\right]\,,
\label{eq:twist2ops}
\end{align}
with subleading contributions from terms involving derivatives that are suppressed in the power-counting, as well as from additional two-body terms that are not Wigner SU(4) symmetric and are suppressed in the large $N_c$ limit. The subscripts 0 and 3 denotes the isoscalar and isovector combinations. Note that the isoscalar contributions arise from matching to both quark and gluon matrix elements (which mix under QCD renormalisation), and will give rise to finite-volume effects that are the same in both cases.

\section{Stochastic variational method in a periodic cubic volume}
\label{sec:svm}

In order to address finite-volume effects in few-nucleon systems in EFT, few-body wavefunctions must be determined subject to the EFT interactions and the given boundary conditions. There are multiple many-body approaches that could be pursued for this task. Two approaches that have been successfully applied are solving the 3-dimensional finite-volume Schr\"odinger equation via discretisation~\cite{Beane:2012ey}, and the stochastic variational method (SVM) for two and three-body systems~\cite{Eliyahu:2019nkz}. The former approach works well for two-body interactions and was effectively used in Ref.~\cite{Beane:2012ey} to analyse hyperon-nucleon interactions where the effective range was not small compared to the spatial extent of the finite volume, $L$, and as such the more direct L\"uscher method \cite{Luscher:1986pf} could not be applied. However, this coordinate-space--based approach scales poorly to larger systems.

The SVM was introduced in nuclear physics~\cite{Varga:1995dm} as a way to sample the  possible spatial, spin, and isospin-wavefunctions for an $A$-nucleon system in a space that is impractically large for an exhaustive approach, see Refs.~\cite{Suzuki:1998bn,Mitroy:2013eom} for reviews. This approach, detailed below (and applied to EFT$_{\pislash}$ in Ref.~\cite{Lensky:2016djr}, for example), involves the construction of a wavefunction by sequential proposals of new stochastically-generated terms and the optimisation of the linear coefficients of the terms by solving the generalised eigenvalue problem of the variational method.
The SVM for a finite volume was first introduced in Ref.~\cite{PhysRevA.87.063609} where systems of bosons in periodic cubic potentials were considered. Periodicity is imposed on the wavefunctions by considering all periodic copies of the infinite-volume potential. The method was first used for nuclei in Ref.~\cite{Eliyahu:2019nkz}, and a similar approach is used here.

\subsection{Finite volume Hamiltonian}

The $n$-particle non-relativistic Hamiltonian that corresponds to the EFT$_{\pislash}$ Lagrangian of Eq.~\eqref{eq:pionlessL} is
\begin{equation}
H=-\frac{1}{2 M_N} \sum_{i} \nabla_{i}^{2}+\sum_{i<j} V_{2}\left({\bf r}_{i j}\right)+\sum_{i<j<k} V_{3}\left({\bf r}_{i j}, {\bf r}_{j k}\right),
\label{eq:Hamiltonian}
\end{equation}
where $i,j,k\in \{1,\ldots, n\}$ label the particle, ${\bf r}_{ij} ={\bf r}_i - {\bf r}_j$ is the displacement between particles $i$ and $j$, and $\nabla^2_i$ denotes the Laplacian for particle $i$. The two and three-particle potentials are
\begin{align}
V_{2}\left({\bf r}_{i j}\right)=&\left(C_{0}+C_{1} \sigma^{(i)} \cdot \sigma^{(j)}\right) g_{\Lambda}\left({\bf r}_{i j}\right)\\
\intertext{and}
V_{3}\left({\bf r}_{i j}, {\bf r}_{j k}\right)={}&D_0 \sum_{c y c} g_{\Lambda}\left({\bf r}_{i j}\right) g_{\Lambda}\left({\bf r}_{j k}\right),
\end{align}
where the interactions have been regulated using Gaussian smearing. This smearing function is given by
\begin{align}\nonumber
g_{\Lambda}({\bf r})={}&\frac{\Lambda^{3}}{8 \pi^{3 / 2}} \exp \left(-\Lambda^{2} |{\bf r}|^{2} / 4\right)\\
={}&\frac{\Lambda^{3}}{8 \pi^{3 / 2}} \prod_{\alpha\in\{x,y,z\}} \exp \left(-\Lambda^{2} r^{(\alpha)2} / 4\right),
\label{eq:regulator}
\end{align}
where ${\bf r}=(r^{(x)},r^{(y)},r^{(z)})$, and is dependant on the regulator parameter $\Lambda$ (also commonly expressed in terms of a length-scale $r_0$, related as $\Lambda=\sqrt{2}/r_0$).
In a finite volume, periodicity can be imposed by replacing $g_\Lambda({\bf r})$ by a regulator which is periodic in each of the spatial directions:
\begin{align}
    g_\Lambda({\bf r},L) =&\frac{\Lambda^{3}}{8 \pi^{3 / 2}}  \prod_{\alpha\in\{x,y,z\}} \nonumber\\
    &{}\times\sum_{q^{(\alpha)}=-\infty}^\infty \exp \left(-\Lambda^{2} (r^{(\alpha)}-L q^{(\alpha)})^{2} / 4\right),
\label{eq:periodic_regulator}
\end{align}
in which the sums run over all periodic copies of the finite volume.

\subsection{Wavefunction ansatz in a finite volume}

While in many applications of the SVM to nuclear systems the angular momentum structure of the wavefunction is tied to the spatial structure due to orbital motion, in a cubic box orbital angular momentum is not a well-defined quantum number. As will be discussed below, one approach in this context is to  build wavefunctions with particular transformation properties under the cubic group, again coupling spatial and spin degrees of freedom. However, since there are a finite number of irreducible representations of the cubic group,  a simpler approach is to consider a factorisation of the spatial and spin-isospin wavefunctions. 

In this work, a trial wave function, $\Psi^{(N)}_h$, is built from linear combinations of symmetrised spatial wavefunctions $\Psi^{{\rm sym}}_{L}$ which satisfy periodic boundary conditions, coupled to the appropriate spin-flavour wavefunction $|\chi_h\rangle$ , i.e.,
\begin{equation}
    \Psi^{(N)}_h\left({\bf x}\right) =\sum_{j=1}^N c_j\Psi^{{\rm sym}}_{L}\left(A_j,B_j,{\bf d}_j; {\bf x}\right) |\chi_h\rangle,
    \label{eq:trialwf}
\end{equation}
where the superscript $(N)$ denotes the total number of terms in the wavefunction, and the dependence of the spatial wavefunction on $h$ is suppressed. The coordinate ${\bf x}=({\bf r}_1,\ldots,{\bf r}_n)$ collects the spatial coordinates of the $n$ particles with ${\bf x}_j={\bf r}_j$. The spatial wavefunctions $\Psi^{{\rm sym}}_{L}$ are detailed in Sec.~\ref{subsec:spatialwf}; the $c_j$, $j\in\{1,\ldots,N\}$, are coefficients, and the  $A_j$, $B_j$ and ${\bf d}_j$ are the parameters of the $j$th spatial wavefunction included in the sum. 
The spin-flavour wavefunction $|\chi_h\rangle$ is a vector\footnote{Note that $\chi$ is common to all terms in Eq.~\eqref{eq:trialwf} in the current implementation. In other approaches for larger systems than will be considered here, $\chi$ is also part of the stochastic sampling and would be indexed by $j$~\cite{Suzuki:1998bn}.} in spin-flavour space for the given nucleus $h$; the particular spin-flavour wavefunctions that are used in this work are given in Sec.~\ref{subsec:spinisospin}.

\subsubsection{Shifted correlated Gaussian spatial wavefunctions}
\label{subsec:spatialwf}

To account for the anisotropy in the spatial wavefunction due to the boundary conditions in a cubic volume, a shifted correlated-Gaussian basis for the trial wavefunctions is used following the approach introduced in Ref.~\cite{PhysRevA.87.063609}.  States are constructed to be antisymmetric under interchange of the spin-flavour degrees of freedom of pairs of nucleons and thus must have symmetric spatial wavefunctions under particle interchange. 

The basic Gaussian structure underlying these wave functions is
\begin{align}    \nonumber 
    \Psi_\infty^{(\alpha)}(A^{(\alpha)},B^{(\alpha)},{\bf d}^{(\alpha)}; {\bf x}^{(\alpha)})&=\exp \left[-\frac{1}{2} \mathbf{x}^{(\alpha)T} A^{(\alpha)} \mathbf{x}^{(\alpha)}\right.
   \\
    &\hspace*{-2.5cm}
  \left. -\frac{1}{2}(\mathbf{x}^{(\alpha)}-\mathbf{d}^{(\alpha)})^{T} B^{(\alpha)}(\mathbf{x}^{(\alpha)}-\mathbf{d}^{(\alpha)})\right]\,,
\end{align}
where ${\bf x}^{(\alpha)}$ is an $n$-component vector collecting the $\alpha$th Cartesian component of the position of each particle.
The $n\times n$ matrices $A^{(\alpha)}$ and $B^{(\alpha)}$,  and $n$-component vector ${\bf d}^{(\alpha)}$, contain the parameters defining the wavefunction. The matrices  $A^{(\alpha)}$ are symmetric, containing $n(n-1)/2$ real parameters, while $B^{(\alpha)}$ are diagonal matrices with $n$ real parameters. 
The finite-volume approach introduced by Yin and Blume \cite{PhysRevA.87.063609} is 
implemented through sums of periodic copies of the intrinsic wavefunction over shifted volumes to define a finite-volume wavefunction 
\begin{equation}
    \Psi_{L}\left(A,B,{\bf d}; {\bf x}\right) = \prod_{\alpha\in\{x,y,z\}} 
    \Psi^{(\alpha)}_{L}\left(A^{(\alpha)},B^{(\alpha)},{\bf d}^{(\alpha)}; {\bf x}^{(\alpha)}\right),
\end{equation}
where $A={\rm diag}\{A^{(x)},A^{(y)},A^{(z)}\}$ is a block-diagonal $3n\times 3n$ matrix that combines the $A^{(\alpha)}$ matrices for each direction, and similarly  $B$ and ${\bf d}$ combine the $B^{(\alpha)}$ and ${\bf d}^{(\alpha)}$ for each direction. The wavefunction for the $\alpha$th direction is 
\begin{align}    \nonumber
    \Psi^{(\alpha)}_{L}\left(A^{(\alpha)},B^{(\alpha)},{\bf d}^{(\alpha)}; {\bf x}^{(\alpha)}\right) &= 
    \\
 & \hspace*{-3cm}   \sum_{{\bf b}^{(\alpha)}} \Psi_\infty^{(\alpha)}(A^{(\alpha)},B^{(\alpha)},{\bf d}^{(\alpha)}; {\bf x}^{(\alpha)}- {\bf b}^{(\alpha)}L),
\end{align}
where the $n$-component vector ${\bf b}^{(\alpha)}$ has components $b^{(\alpha)}_j\in \mathbb{Z}$.
The resulting wavefunction $\Psi_{L}\left(A,B,{\bf d}; {\bf x}\right)$ satisfies the periodic constraint
\begin{equation}
\Psi_{L}\left(A,B,{\bf d}; {\bf x}\right)=\Psi_{L}\left(A,B,{\bf d}; {\bf x} + {\bf n} L\right)
\end{equation}
for all integer $3n$-tuples\footnote{In practice, these sums are truncated as discussed in Appendix \ref{app:MEintegrals}.}, $\mathbf{n}\in \mathbb{Z}^{3n}$.
In order to symmetrise the wavefunction under particle exchange, the rows and columns of each $A^{(\alpha)}$ and $B^{(\alpha)}$  and rows of ${\bf d}^{(\alpha)}$ are interchanged under all $n!$ possible permutations, ${\cal P}$, of particles. That is
\begin{equation}
    \Psi^{{\rm sym}}_{L}\left(A,B,{\bf d}; {\bf x}\right) = \sum_{\cal P} \Psi_{L}\left(A_{\cal P},B_{\cal P},{\bf d}_{\cal P}; {\bf x}\right),
    \label{eq:symwf}
\end{equation}
where $A_{\cal P}$ is the permuted form of $A$ and similarly for $B_{\cal P}$ and ${\bf d}_{\cal P}$. 
As discussed in Appendix \ref{app:scattering}, the 
shifted Gaussian basis is able to describe scattering states at finite volume as well as compact bound states.

\subsubsection{Cubic harmonics}

A periodic spatial volume with identical extent in each direction has an underlying cubic symmetry and is invariant under action of elements of the cubic group, $H_3$.
By imposing cubic symmetry, wavefunctions that transform in particular representations of  $H_3$ can be constructed, potentially allowing more efficient exploration of the space of correlated shifted Gaussians. Each term in the variational wavefunction can be constructed to respect the given transformation properties rather than relying on stochastic sampling of a sum of terms to discover the symmetry approximately.
The $H_3$-covariant wavefunction transforming in the representation ${R}$ of $H_3$ is 
given by
\begin{equation}
    \Psi^{{\rm sym},R}_{L}\left(A,B,{\bf d}; {\bf x}\right) = \sum_{p} c^{(R)}_p\Psi^{{\rm sym}}_{L}\left(A_p,B_p,{\bf d}_{p}; {\bf x}_p\right),
    \label{eq:H3improvedwf}
\end{equation}
where $p$ indexes the permutations of the Cartesian directions, $c_p^{(R)}$ are constants defining the representation, and $A_p$ is the appropriately block-permuted form of the matrix $A$ and similarly for $B_p$ and ${\bf d}_p$. For the ground states that are considered here, the $A_1$ (trivial) representation of $H_3$ is assumed, for which $c^{(A_1)}_p=1$.
The utility of using Eq.~\eqref{eq:H3improvedwf} instead of  Eq.~\eqref{eq:symwf}, has been investigated. Overall, it is found that $N$-term wavefunctions constructed from terms of the form $\Psi^{{\rm sym},R}_{L}$ are about a factor of  five  better approximations than $N$-term wavefunctions constructed from terms from Eq.~\eqref{eq:symwf}, measured in terms of the number of wavefunction terms required to achieve convergence within a given tolerance. However the cost of evaluation of the matrix elements needed in the SVM is a factor of six slower using Eq.~\eqref{eq:H3improvedwf} than using Eq.~\eqref{eq:symwf}. In the primary studies of this work, trial wavefunctions are thus constructed using the simpler ansatz in Eq.~\eqref{eq:symwf}.

\subsubsection{Spin-flavour wavefunctions}
\label{subsec:spinisospin}

The simplest spin-flavour wavefunctions for the small nuclei that are considered in this work are straightforward to construct explicitly.
In particular, the necessary states are defined as
\begin{align}\nonumber
    \left|\chi_{d,j_z=+1}\right\rangle =& \frac{1}{\sqrt{2}}\left[
    \ket{p^\uparrow n^\uparrow} -  \ket{n^\uparrow p^\uparrow} 
    \right]\,,
        \\\nonumber
    \left|\chi_{d,j_z=0}\right\rangle =& \frac{1}{2}\left[
    \ket{p^\uparrow n^\downarrow} - \ket{n^\uparrow p^\downarrow} 
    + \ket{p^\downarrow n^\uparrow} -  \ket{n^\downarrow p^\uparrow} 
    \right]\,,
    \\\nonumber
    \left|\chi_{pp}\right\rangle =& \frac{1}{\sqrt{2}}\left[
    \ket{p^\uparrow p^\downarrow} -  \ket{p^\downarrow p^\uparrow} 
    \right]\,,
    \\\nonumber
    \left|\chi_{np,j=0}\right\rangle =& \frac{1}{2}\left[
    \ket{p^\uparrow n^\downarrow} + \ket{n^\uparrow p^\downarrow} 
    - \ket{p^\downarrow n^\uparrow} -  \ket{n^\downarrow p^\uparrow} 
    \right]\,,
    \\\nonumber
    \left|\chi_{\hthree,j_z=1/2}\right\rangle =& \frac{1}{\sqrt{6}}\left[
    \ket{n^\uparrow p^\uparrow n^\downarrow} - \ket{ n^\downarrow p^\uparrow n^\uparrow}
     -\ket{p^\uparrow n^\uparrow n^\downarrow}\right. \nonumber \\ \nonumber & \left. + \ket{ p^\uparrow n^\downarrow n^\uparrow}
    - \ket{n^\uparrow n^\downarrow p^\uparrow } + \ket{ n^\downarrow n^\uparrow p^\uparrow}
    \right]\,,
    \\\nonumber
    \left|\chi_{\hethree,j_z=1/2}\right\rangle =& \frac{1}{\sqrt{6}}\left[
    \ket{p^\uparrow n^\uparrow p^\downarrow} - \ket{ p^\downarrow n^\uparrow p^\uparrow}
     -\ket{n^\uparrow p^\uparrow p^\downarrow}\right.  \\ 
     & \left. + \ket{ n^\uparrow p^\downarrow p^\uparrow}
    - \ket{p^\uparrow p^\downarrow n^\uparrow } + \ket{ p^\downarrow p^\uparrow n^\uparrow}
    \right]\,,
\end{align}
where $p^{\uparrow(\downarrow)}$ and $n^{\uparrow(\downarrow)}$ denote proton and neutron states of the given spin.

\subsection{Implementation of the stochastic variational method}
\label{sec:implmentation}

The trial wavefunction $\Psi^{(N)}$ of Eq.~\eqref{eq:trialwf} is constructed so as to minimize the bound which it provides on the ground-state energy:
\begin{equation}
    E^h_0 \leq  \frac{\int\Psi^{(N)}_h({\bf x})^\ast H \Psi^{(N)}_h({\bf x}) d{\bf x}}{\int\Psi^{(N)}_h({\bf x})^\ast\Psi^{(N)}_h({\bf x}) d{\bf x}}.
    \label{eq:variational}
\end{equation}
Because of the Gaussian structure of the trial wavefunction, the various contributions to the Hamiltonian matrix element, which are $3n$-dimensional integrals, can be evaluated analytically \cite{PhysRevA.87.063609}, as shown in Appendix \ref{app:MEintegrals}.

In the current application of the SVM, $\Psi^{(N)}$ is built up from 1 to $N$ terms via an iterative procedure as follows:
\begin{enumerate}
    \item Given an $M$-term wavefunction (where $M<N$) defined by matrices $A_j$, $B_j$, ${\bf d}_j$ for $j\in\{1,\ldots,M\}$, $N_{\rm proposal}$ proposed candidates for the $(M+1)$th  term are constructed by randomly choosing matrices $A_{M+1}$, $B_{M+1}$ and ${\bf d}_{M+1}$. For simplicity of notation, the spatial wavefunction of the $j$th term is denoted as $\Psi_j({\bf x}) \equiv \Psi^{{\rm sym}}_{L}\left(A_j,B_j,{\bf d}_j; {\bf x}\right)$.
    
    \item For each candidate term $\Psi_{M+1}(\bf{x})$, the normalisation integrals 
    \begin{align}\label{eq:Nint}
        [\mathbb{N}^{(M+1)}]_{ij}\equiv& \int \Psi_i\left({\bf x}\right)^\ast 
        \Psi_j\left({\bf x}\right) d{\bf x}
    \end{align}
    and Hamiltonian matrix elements 
    \begin{align}\label{eq:Hint}
        [\mathbb{H}^{(M+1)}]_{ij}\equiv&\int \Psi_i\left({\bf x}\right)^\ast 
        \langle \chi_h| H  |\chi_h\rangle \Psi_j\left( {\bf x}\right) d{\bf x} 
    \end{align}  
    are computed for $\{i,j\}\in\{1,\ldots,M+1\}$ and used to define matrices $\mathbb{N}^{(M+1)}$ and $\mathbb{H}^{(M+1)}$, respectively. Note that only the additional $(M+1)$th row and column must be computed, given that the $\mathbb{N}^{(M)}$ and $\mathbb{H}^{(M)}$ matrices were computed in the previous iteration.
    
    \item The generalised eigenvalue problem 
    \begin{equation}
        \mathbb{H}^{(M+1)} \,\mathbf{c} = \lambda\, \mathbb{N}^{(M+1)} \, \mathbf{c} 
        \label{eq:eval}
    \end{equation}
    is solved for the eigenvalues $\lambda_0^{(M+1)}\le \lambda_1^{(M+1)} \le\ldots\le  \lambda_{M+1}^{(M+1)}$, and coefficient vectors $\mathbf{c}_\ell^{(M+1)}=(c_1,\ldots, c_{M+1})$ for $\ell\in\{1,\ldots, M+1\}$ labelling the eigenvalue. 
    
    \item The  wavefunction which results in the smallest eigenvalue $\lambda_0^{(M+1)}$ is selected from the set of $N_{\rm proposal}$ candidates\footnote{Note that each of the $N_{\rm proposal}$ lowest eigenvalues $\lambda_0^{(M+1)}$ is smaller than the lowest eigenvalue from the previous iteration $\hat\lambda_0^{(M)}$.} and added to the iteratively-constructed trial wavefunction to define $\Psi^{(M+1)}\left({\bf x}\right)$. 
\end{enumerate}
Naturally, the optimization at each step of this iterative procedure depends on the Hamiltonian $H$ and hence on the LECs that define it; to enable optimization of the trial wavefunction across a broad range of LECs, in the numerical study undertaken here the values of the LECs are varied for each step of optimization, cycling through $N_{\rm couplings}$ choices that span the relevant parameter spaces. After initializing the procedure with the first trial wavefunction (the $M=1$ term), for which the matrices $\mathbb{N}^{(1)}$ and $\mathbb{H}^{(1)}$ are single numbers and the generalised eigenvalue problem is trivial, additional terms are added iteratively until the wavefunction has a fixed number of terms, $N$. $N$ must be taken large enough that the optimization procedure has converged by some definition. Here, $N$ is set by the criterion that repeated optimizations based on different random seeds achieve the same minimum energies within some tolerance, and that adding some fixed number of additional terms to the trial wavefunctions does not alter the minimum found, within the same tolerance. Details of the optimisation procedure for the cases considered in this study are provided in Sec.~\ref{sec:groundstatematching}.

\section{Ground states of two and three-nucleon systems}
\label{sec:groundstatematching}

In this work, the finite-volume SVM is used to match the LECs of FVEFT$_{\pislash}$ to the LQCD results for two and three-nucleon systems which were obtained in Ref.~\cite{Beane:2012ey}, where nuclear states with SU(3)$_f$-symmetric quark masses corresponding to $m_\pi=806$ MeV were studied in three volumes of spatial extents $L\in\{3.4,\, 4.5,\,6.7\}$~fm. The binding energies extracted in that work, defined as $\Delta E_h=E_h-A E_p$ where $A$ is the atomic number of the state $h$, are tabulated in Table \ref{tab:bindings}. This matching procedure, used in this work as a first step in the study of the matrix elements of various currents in these states in this framework, closely mirrors, and reproduces, the analysis of Ref.~\cite{Eliyahu:2019nkz}.
\begin{table}[!t]
    \centering
\begin{tabular}{cccc}
\hline \hline $h$ & $L=3.4$ fm & $L=4.5$ fm & $L=6.7$ fm \\
\hline $pp$ & $17.8(3.3)$ & $15.1(2.8)$ & $15.9(3.8)$ \\
$d$ & $25.4(5.4)$ & $22.5(3.5)$ & $19.5(4.8)$ \\
${ }^{3} \mathrm{H}$ & $65.6(6.8)$ & $63.2(8.0)$ & $53.9(10.7)$ \\
\hline \hline
\end{tabular}
    \caption{Finite-volume binding energies [MeV] determined in the LQCD calculations of Ref.~\cite{Beane:2012ey}.}
    \label{tab:bindings}
\end{table}

\subsection{Two-body states}
\label{sec:two}

\begin{figure}[!t]
\begin{center}
\includegraphics[width=0.95\columnwidth]{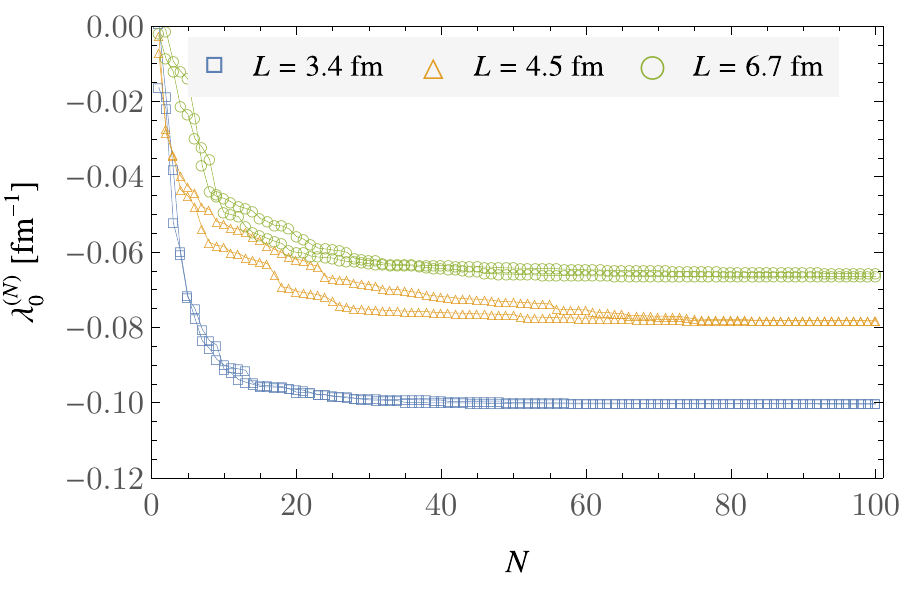}
	\caption{\label{fig:2bodyconvergence}Convergence of the eigenvalues $\lambda_0^{(N)}$ to the ground-state energy of the diproton system as additional terms are added to the variational wavefunction. The three colours correspond to wavefunctions optimized for the $L=3.4$~fm (blue), 4.5~fm (orange), and 6.7~fm (green) volumes, with the LEC $C_T$ set to its optimised value after matching to the LQCD results, as discussed in the main text. The results are shown for one example of the regulator parameter corresponding to $r_0=0.2$~fm. For each volume, the results of optimisation procedures starting with two random seeds are shown.}
\end{center} 
\end{figure}

\begin{figure}[!t]
\includegraphics[width=0.9\columnwidth]{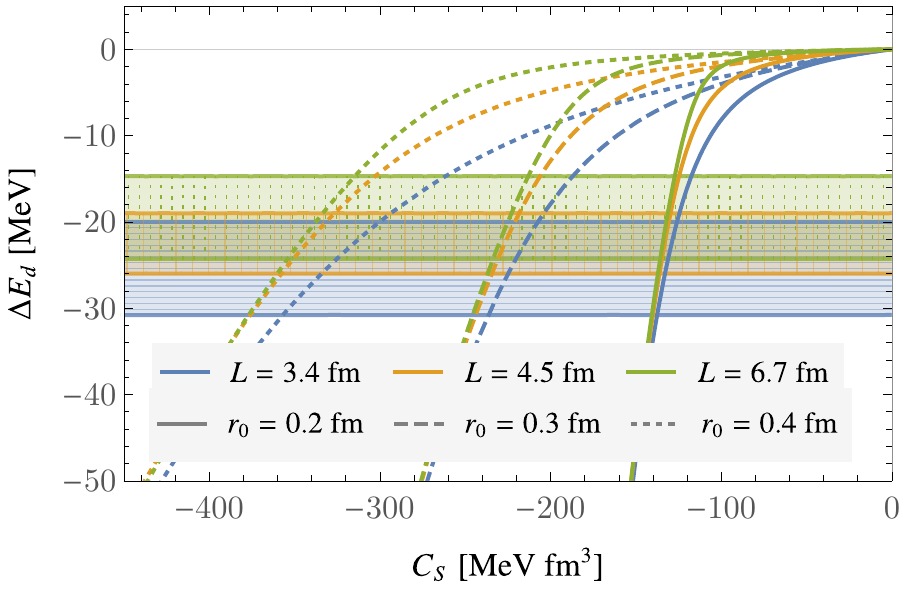}\\[15pt]
\includegraphics[width=0.9\columnwidth]{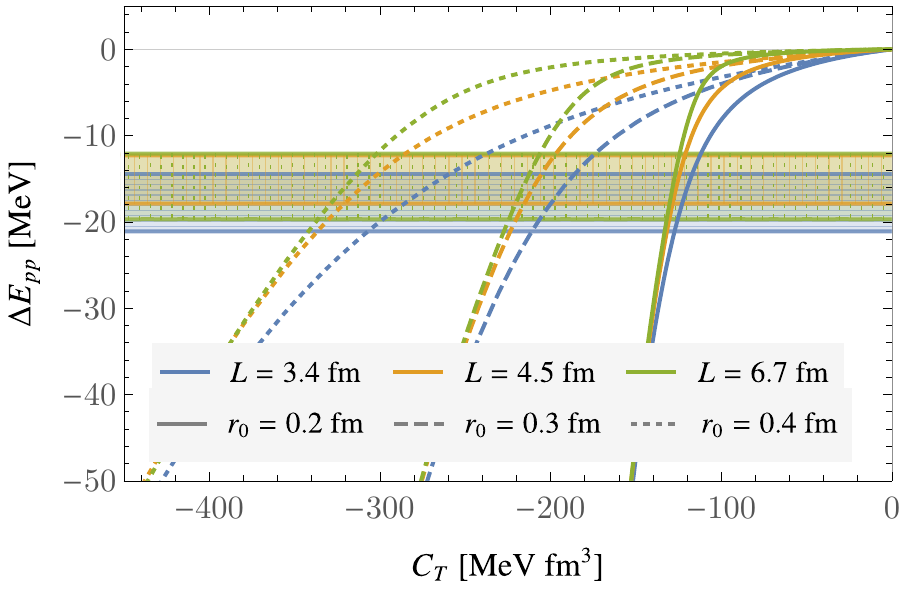}
	\caption{\label{fig:2bodyvsC} Binding energies of the deuteron (upper panel) and diproton (lower panel) systems as a function of the relevant two-body EFT$_{\pislash}$ LECs. The three sets of curves correspond to the three different choices of the regulator scale $\Lambda=\sqrt{2}/r_0$ corresponding to $r_0\in\{0.2,0.3,0.4\}$~fm (solid, dashed, and dotted) and the three colours correspond to the three different volumes: $L=3.4$~fm (blue), 4.5~fm (orange), and 6.7~fm (green). The horizontal bands show the values of the binding energies for each volume from the LQCD calculations of Ref.~\cite{Beane:2012ey}.}
\end{figure}

For each of the three finite volumes for which Ref.~\cite{Beane:2012ey} provides LQCD data, variational wavefunctions were optimised first for the $pp$ and $d$ two-body systems. 
As described in  Sec.~\ref{sec:implmentation}, the undetermined coefficient of the two-body potential in the Hamiltonian (which corresponds to the LEC $C_S$ in the case of the deuteron and $C_T$ in the case of $pp$) is varied over $N_{\rm couplings}=10$ different choices throughout the optimisation procedure, which are chosen to be evenly-spaced corresponding approximately to the plot range of Fig.~\ref{fig:2bodyvsC}; both two-body systems are thus optimised simultaneously.
Taking $N_\text{proposal}=30$, it is found for all two-body optimizations undertaken in this work---at each finite volume, and for each of three choices of the regulator $\Lambda=\sqrt{2}/r_0$ corresponding to $r_0\in\{0.2,0.3,0.4\}$~fm---that after 100 terms have been added to the wavefunction, the last 20 terms are within 1\% (typically, within 0.1\%) of the final value for each of the values of the coupling that are used in optimization, and also that optimizations starting with different random seeds agree within that same tolerance. An example of this convergence is shown in Fig.~\ref{fig:2bodyconvergence}.

\begin{figure}[!t]
\includegraphics[width=0.9\columnwidth]{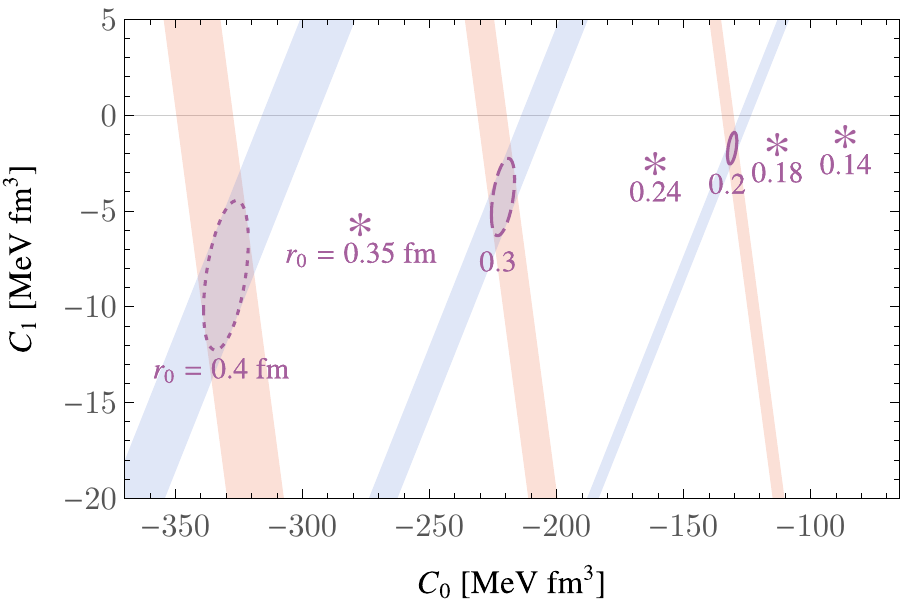}
	\caption{\label{fig:c0vsc1} The EFT$_{\pislash}$ couplings $C_{0,1}$ determined by fitting to the results of the LQCD calculations. The three sets of intersecting bands (blue for $pp$ and red for $d$) and corresponding ellipses show results obtained with wavefunctions optimised with the three different values of the regulator scale $\Lambda=\sqrt{2}/r_0$ studied here. The  asterisks show the results of an analogous analysis undertaken in Ref.~\cite{Eliyahu:2019nkz}, with different values of the regulator $r_0$, as indicated on the figure.}
\end{figure}

The ground-state $d$ and $pp$ binding energies resulting from the optimised variational wavefunction are shown as a function of the relevant undetermined LEC in Fig.~\ref{fig:2bodyvsC}, for all three finite volumes and for all choices of the regulator parameter $\Lambda$ which are studied. Comparing with the LQCD results for the binding energies in each lattice volume, it is clear that for each value of $\Lambda$ the results in all volumes are consistent with a single consistent value of the relevant coupling, indicating that there is no need to introduce higher order terms in EFT$_{\pislash}$. The best-fit values of the corresponding couplings, which depend on the regulator scale, are determined through a combined fit to the three volumes and are presented in Table~\ref{tab:couplings} and Fig.~\ref{fig:c0vsc1}.
Note that the EFT$_{\pislash}$ interaction proportional to $C_1$ is suppressed by $1/N_c^2$ relative to the Wigner-symmetric interaction with coefficient $C_0$ in the large-$N_c$ limit; this hierarchy is born out in the fitted values of the couplings.
\begin{table}[!t]
    \centering
\begin{tabular}{cccc}
\hline \hline  $r_0$ [fm]& $0.2$ & $0.3$ & $0.4$  \\
\hline  
$C_0$ & $-131(2)$ & $-220(5)$ & $-330(9)$ \\
$C_1$ & $-2(1)$ & $-4(2)$ & $-8(4)$ \\ \hline
$C_S$ & $-133(2)$ & $-225(6)$ & $-338(11)$ \\
$C_T$ & $-126(2)$ & $-208(6)$ & $-305(11)$ \\ \hline
$D_0$ & $17(2)$ & -- & -- \\
\hline \hline
\end{tabular}
    \caption{The two and three-body EFT$_{\pislash}$ LECs determined from matching the SVM calculations to the LQCD energy shifts for each value of the cutoff parameter $r_0$. For the two-body case, $C_S$ and $C_T$ are determined in terms of $C_{0,1}$ through Eq.~\eqref{eq:C01CSTrelation}. For $D_0$, three-nucleon optimisations were only performed for $r_0=0.2$~fm.}
    \label{tab:couplings}
\end{table}
  
\begin{figure}[!t]
\includegraphics[width=0.9\columnwidth]{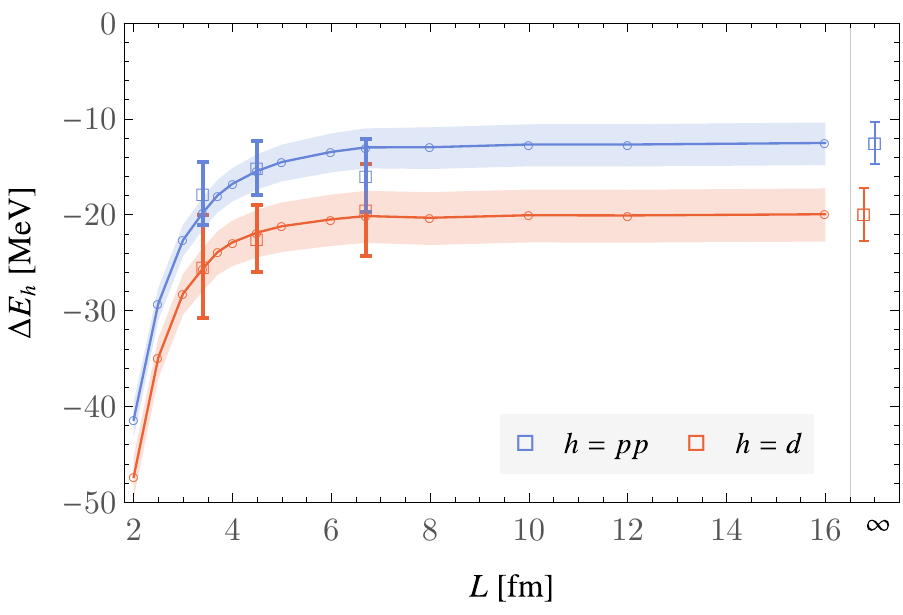}
	\caption{\label{fig:2bidybindingsvsL} The volume-dependence of the deuteron and diproton binding energies compared with the LQCD data which was used to determine the relevant LECs. The infinite-volume extrapolations of the binding energies, computed as described in the text, are shown in the rightmost sub-panel. }
\end{figure}

Having determined the two-body couplings, wavefunctions optimised in infinite volume in exactly the same way are used to determine the binding energies in the infinite-volume limit. Figure~\ref{fig:2bidybindingsvsL} shows the volume-dependence of the binding energies of the deuteron and diproton systems (in order to show a curve, binding energies computed with wavefunctions optimised at additional intermediate volumes are also shown). Extrapolations are shown for $r_0=0.2$ fm, but the extrapolations for other values  are indistinguishable. Although the values of the LECs depend on the value of the regulator $r_0$, the resulting finite and infinite-volume energies are regulator-independent. Table \ref{tab:extrapolated_bindings} lists the extrapolated binding energies  and compares them to Refs.~\cite{Beane:2012vq,Eliyahu:2019nkz}, with which they are in close agreement. 
\begin{table}[!t]
    \centering
\begin{tabular}{cccc}
\hline \hline $h$ & $L=\infty$ & Ref.~\cite{Beane:2012vq} & Ref.~\cite{Eliyahu:2019nkz} \\
\hline $\mathrm{pp}$ & -12.5(2.2) & -15.9(3.8)  & -13.8(1.8)\\
${ }^{2} \mathrm{H}$ & -19.9(2.8) & -19.5(4.8)  & -20.2(2.3)\\
${ }^{3} \mathrm{H}$ &.-60.2(6.5) & -53.9(10.7) & -58.2(4.7)\\
\hline \hline 
\end{tabular}
    \caption{The extrapolated infinite-volume binding energies determined in the SVM approach for the two and three-body systems. Also shown are the binding energies determined in the original LQCD study~\cite{Beane:2012vq} and in Ref.~\cite{Eliyahu:2019nkz} also using the SVM method.}
    \label{tab:extrapolated_bindings}
\end{table}

\subsection{Three-body states}
\label{sec:three}
  
\begin{figure}[!t]
\includegraphics[width=0.9\columnwidth]{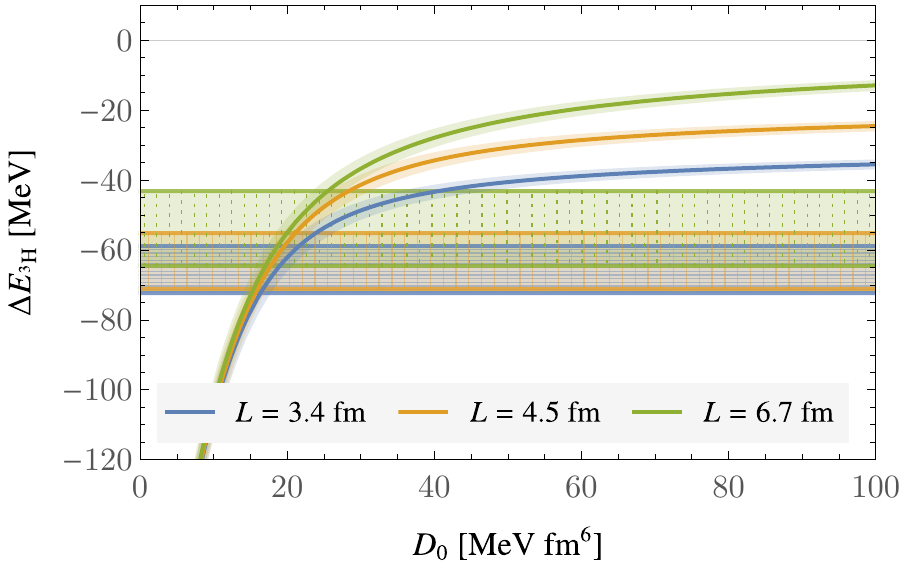}
	\caption{\label{fig:3bodybindingvsD0} The dependence of the triton binding energy on the three-body coupling $D_0$ in each of three finite volumes for which there is LQCD data, $L=3.4$~fm (blue), 4.5~fm (orange), and 6.7~fm (green). The curves are shown for a regulator scale $r_0=0.2$~fm and for the values of the two-body couplings determined in Sec. \ref{sec:two}, with the shading on the bands indicating the uncertainty which arises as these couplings are varied within their uncertainties. Other details are as in Fig.~\ref{fig:2bodyvsC}.}
\end{figure}

Having determined the two-nucleon couplings, an analogous procedure is repeated for the triton to determine the coefficient $D_0$ of the leading-order three-nucleon coupling in Eq.~\eqref{eq:Lagrangian3body}.
The stochastic optimisation of the three-body wavefunction is performed with the two-body LECs fixed to their central values determined as discussed in the previous subsection, for a single value of the Gaussian regulator parameter $r_0=0.2$~fm. As for the two-body case, $N_{\rm couplings}=10$ values of the three-body LEC $D_0$ are cycled through in the wavefunction construction procedure, spanning the relevant coupling range (corresponding approximately to the range of the horizontal axis in Fig.~\ref{fig:3bodybindingvsD0}). The same convergence criteria as in the two-body case are satisfied after wavefunctions with $N=250$ terms have been constructed.

\begin{figure}[!t]
\includegraphics[width=0.9\columnwidth]{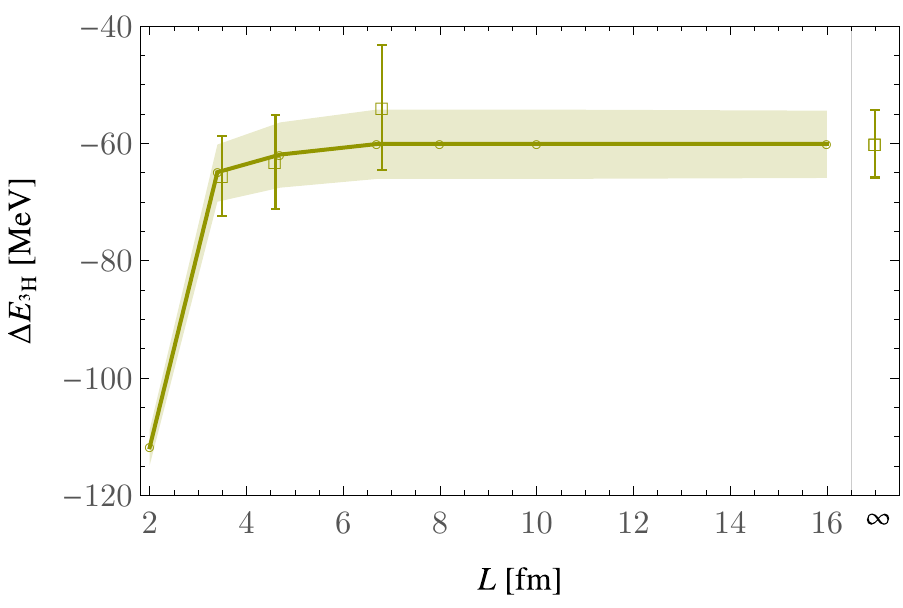}
	\caption{\label{fig:3bodybindingvsL} The volume-dependence of the triton binding energy, compared with the LQCD data. The infinite-volume extrapolation of the binding energy, computed as described in the text, is shown in the rightmost sub-panel. }
\end{figure}

Figure \ref{fig:3bodybindingvsD0} shows the dependence of the triton binding energy on the three-body coupling, $D_0$, for the optimal values of the two-body couplings; the shaded bands around the curves show the result of varying the two-body LECs within their uncertainties for wavefunctions optimised in each of the three volumes for which there is LQCD data.
Figure \ref{fig:3bodybindingvsL} and Table \ref{tab:extrapolated_bindings} show the infinite-volume extrapolation of the triton binding energy. As for the two-body systems, the extrapolations and couplings reported here are in close agreement with those of Ref.~\cite{Eliyahu:2019nkz}.

\section{Matrix elements in the stochastic variational method}
\label{sec:ME}

Having determined finite-volume ground state (or in principle excited state) wavefunctions in the SVM, those wavefunctions can be used to evaluate finite-volume matrix elements of operators in FVEFT$_{\pislash}$.  The transition matrix element between an initial state $\Psi_i$ and final state $\Psi_f$ is given by
\begin{equation}
    \frac{\langle \Psi_f | {\cal O} | \Psi_i\rangle}{\sqrt{\langle \Psi_f | \Psi_f\rangle \langle \Psi_i | \Psi_i\rangle}},
\end{equation}
where ${\cal O}$ is a generic local EFT operator and bra-ket notation is used for concision.\footnote{Matrix elements of non-local products of operators (such as those relevant for double-$\beta$ decay) can also be approached in the finite-volume SVM as discussed in Sec.~\ref{sec:discuss}.}

Here, matrix elements of axial, electromagnetic, scalar, and tensor currents and the unpolarised twist-two operators are studied. In each case, the relevant EFT$_{\pislash}$ currents of Sec.~\ref{sec:pionless} are translated into operators acting on the $n$-body states.
As with the nucleon-nucleon strong interactions, two-body contributions to the various currents are regulated using the Gaussian approach as in Eq.~\eqref{eq:regulator} and rendered periodic using Eq.~\eqref{eq:periodic_regulator}. 
Specifically, each two-body current is implemented as
\begin{align}
\left[\left(N^{\dagger}({\bf r}_i) \Sigma N({\bf r}_i)\right)\left(N({\bf r}_j)^{\dagger} \Sigma^\prime N({\bf r}_j)\right)+\text{h.c.}\right] g_\Lambda({\bf r}_{ij},L)\,,
\end{align}
where $\Sigma^{(\prime)}$ denotes a spin-isospin structure and $g_\Lambda({\bf r},L)$, defined as in Eq.~\eqref{eq:regulator}, implements a periodic regulated form of the $\delta$-function implied in local two-body EFT currents. 
Since the matrix elements that are considered are for zero momentum transfer, the current is integrated over the positions ${\bf r}_{i,j}$.
For each current, $X\in\{A_{i,a}, J_{i}^{EM},  S_a, T_{ij,a}, O^{(n)}_a\}$ for $a\in\{0,1,2,3\}$, the zero--momentum-projected, regulated form is labelled as ${\cal X}$. 

The evaluation of the relevant matrix elements factorises into a spin-isospin calculation that is specific to each type of operator, and a calculation of the  matrix element of the spatial wavefunction. Since all currents that are considered  enter with the spatial dependence determined by the Gaussian regulator function, these latter spatial matrix elements have a common form and are given for diagonal matrix elements in state $h$ by
\begin{equation}
      h_{h}(\Lambda,L)=\frac{\int \prod_kd^3{\bf r}_k \sum_{i< j} g_\Lambda({\bf r}_{ij},L) |\psi_h({\bf R}_n)|^2}{\int \prod_kd^3{\bf r}_k|\psi_h({\bf R}_n)|^2} \,,
\end{equation}
where ${\bf R}_n=\{{\bf r}_1,\ldots, {\bf r}_n\}$ indicates dependence on the coordinates of each of the $n$ particles.
For transition matrix elements between states  $a$ and $b$, the corresponding expression is
\begin{equation}
      h_{a \leftarrow b}(\Lambda,L)=\frac{\int \prod_kd^3{\bf r}_k \sum_{i< j} \psi^\ast_a({\bf R}_n) g_\Lambda({\bf r}_{ij},L) \psi_b({\bf R}_n)}{\sqrt{\int \prod_kd^3{\bf r}_k|\psi_a({\bf R}_n)|^2 \int \prod_kd^3{\bf r}_k|\psi_{b}({\bf R}_n)|^2}} \,.
\end{equation}

The relevant one and two-body LECs of currents in FVEFT$_{\pislash}$ can be tuned such that the EFT$_{\pislash}$ matrix elements determined in this way reproduce the matrix elements (or their ratios to the proton matrix element) determined in LQCD in a particular lattice volume or set of volumes. In what follows, the LQCD matrix elements in atomic number $A=\{2,3\}$ systems calculated  at  $m_\pi=806$ MeV on the $L=4.5$~fm ensemble, discussed above, are used. 
The EFT$_{\pislash}$ counter-terms determined in this way are specified in the Gaussian-regulated scheme and should not be compared with the corresponding counterterms determined in dimensional regularisation. Indeed, for the current purposes, the extraction of counterterms can be viewed simply as an intermediate step in extracting matrix elements at infinite volume. The infinite-volume--extrapolated matrix elements can be matched to EFT$_{\pislash}$ regulated in more common schemes, such as dimensional regularisation or the power-divergent subtraction scheme~\cite{Kaplan:1998sz}, to determine the two-body LECs for comparison to other extractions.

\subsection{Axial matrix elements: proton-proton fusion, tritium $\beta$-decay and isoscalar charges}
\begin{figure*}[p]
	\subfigure[\label{fig:axisovecvsL1A}]{\includegraphics[width=0.85\columnwidth]{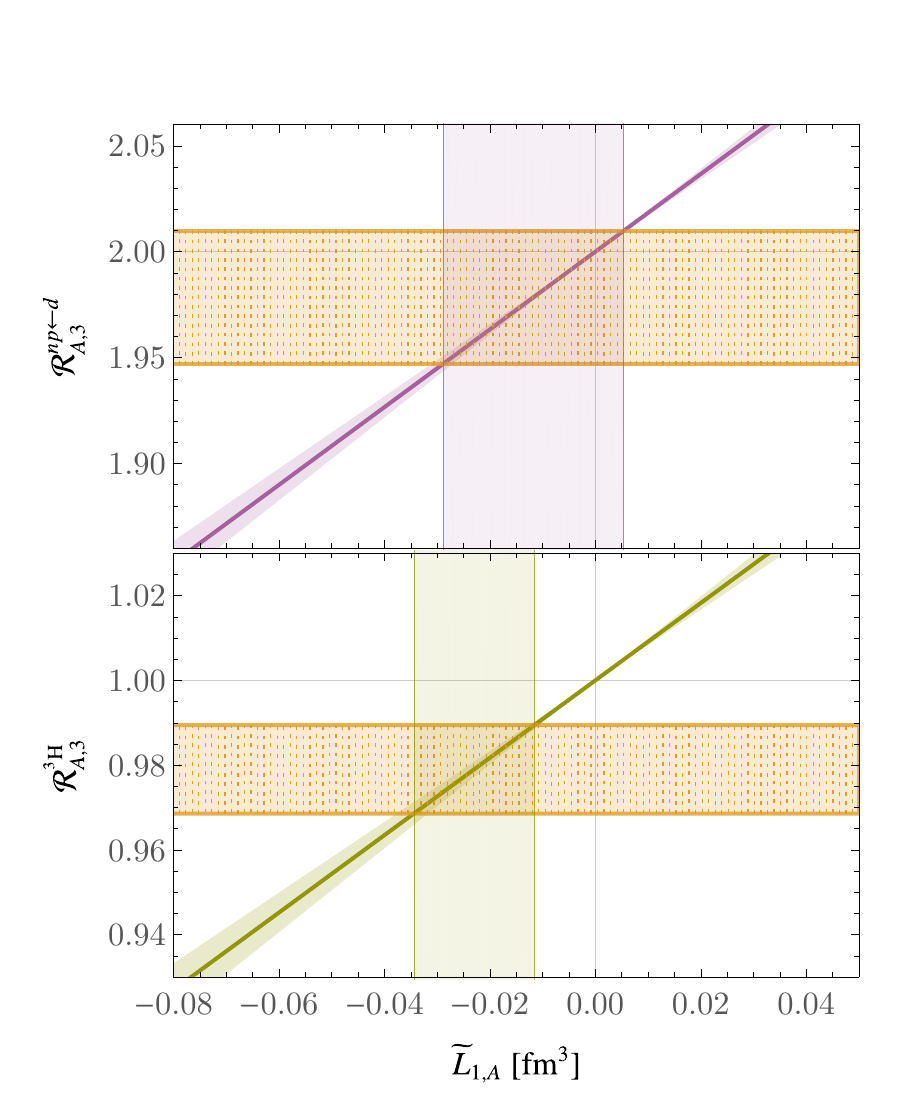}}
	\subfigure[\label{fig:axisovecvsL}]{\includegraphics[width=0.85\columnwidth]{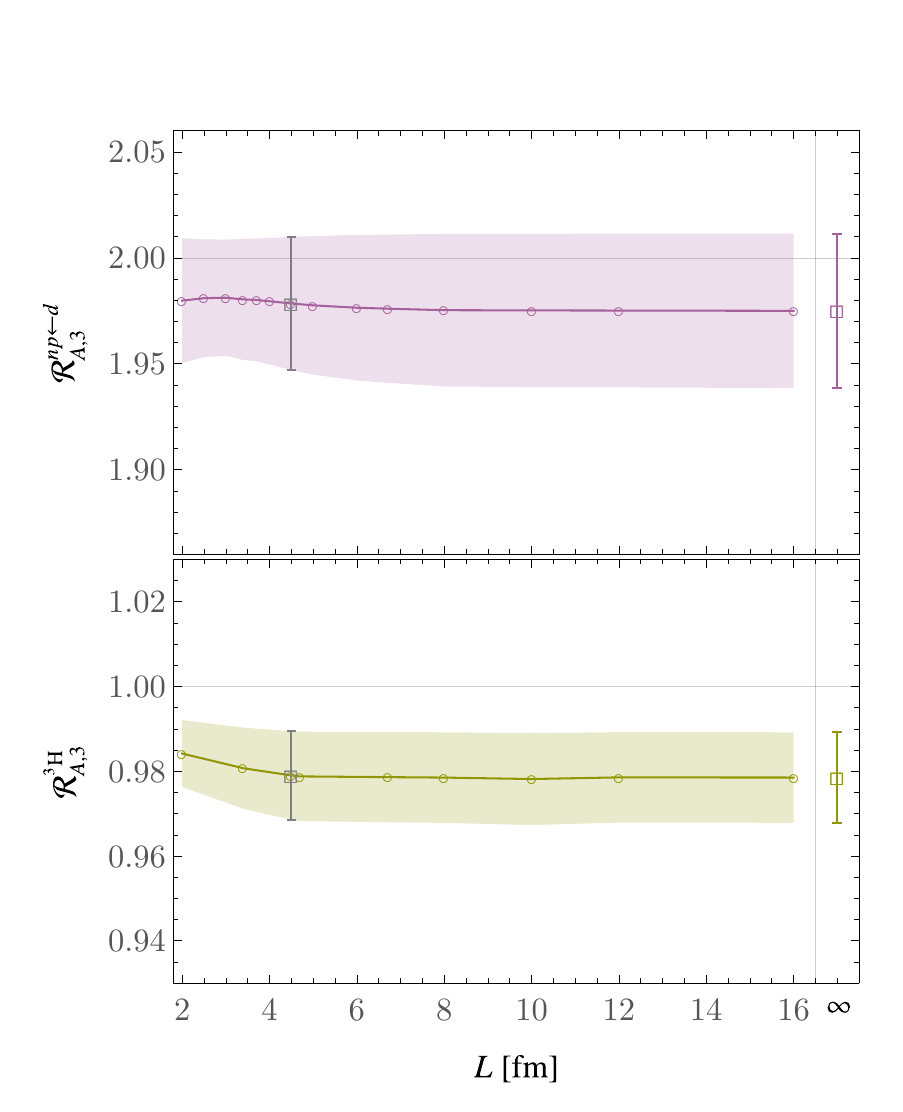}}
	\caption{\label{fig:axialisovec} (a) The dependence of the $d\to np$ (upper) axial transition matrix element and the $\hthree$ (lower) matrix element on the two body axial LEC ratio. The horizontal bands show the constraints from the LQCD calculation of Refs.~\cite{Savage:2016kon,Chang:2017eiq}, with $L=4.5$~fm, and the vertical bands highlight the region of coupling that is consistent at one standard deviation with the LQCD result, for the central values of $C_{0,1}$ and $D_0$ determined in Secs.~\ref{sec:two} and \ref{sec:three}.
	(b) The dependence of the matrix elements on the spatial extent of the lattice, $L$. The LQCD constraint is shown as the grey data point. The infinite-volume limits are shown at the right edge of the figure.}
\vspace{5mm}
    \subfigure[]{\includegraphics[width=0.85\columnwidth]{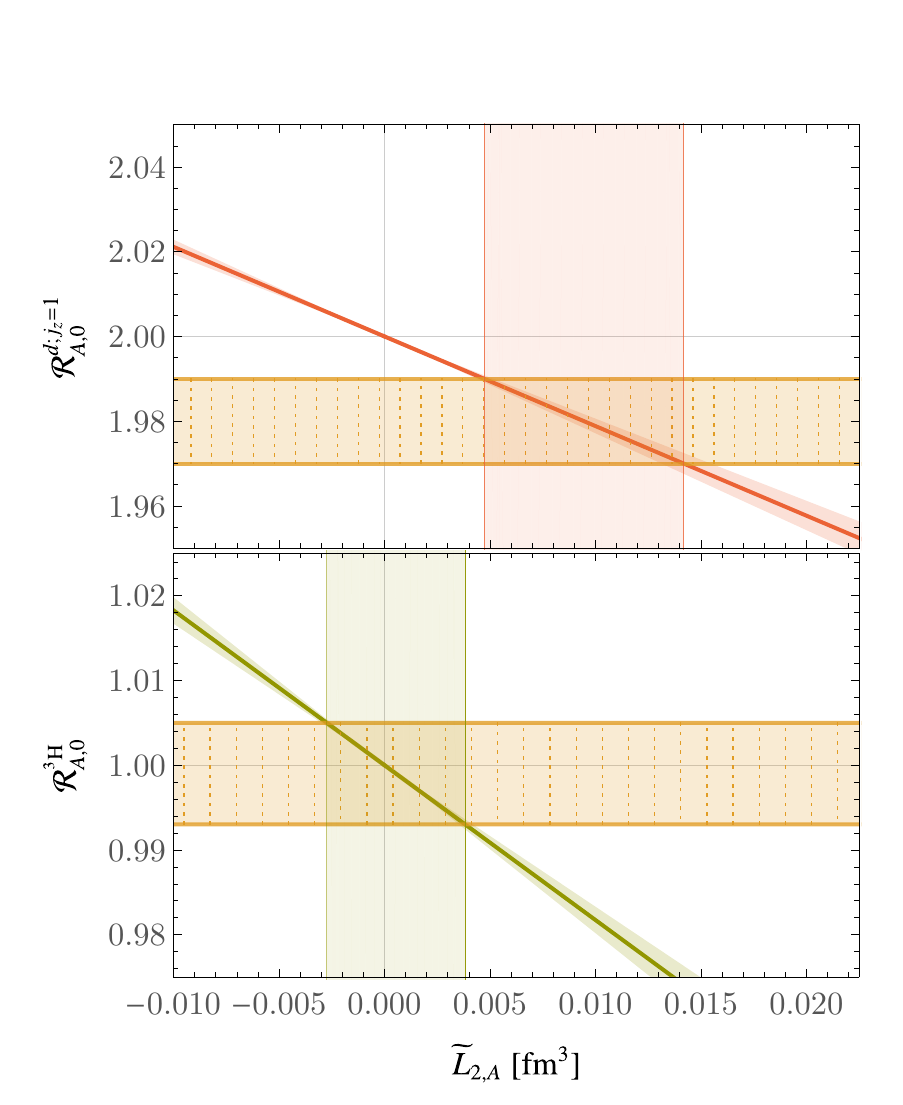}}
	\subfigure[]{\includegraphics[width=0.85\columnwidth]{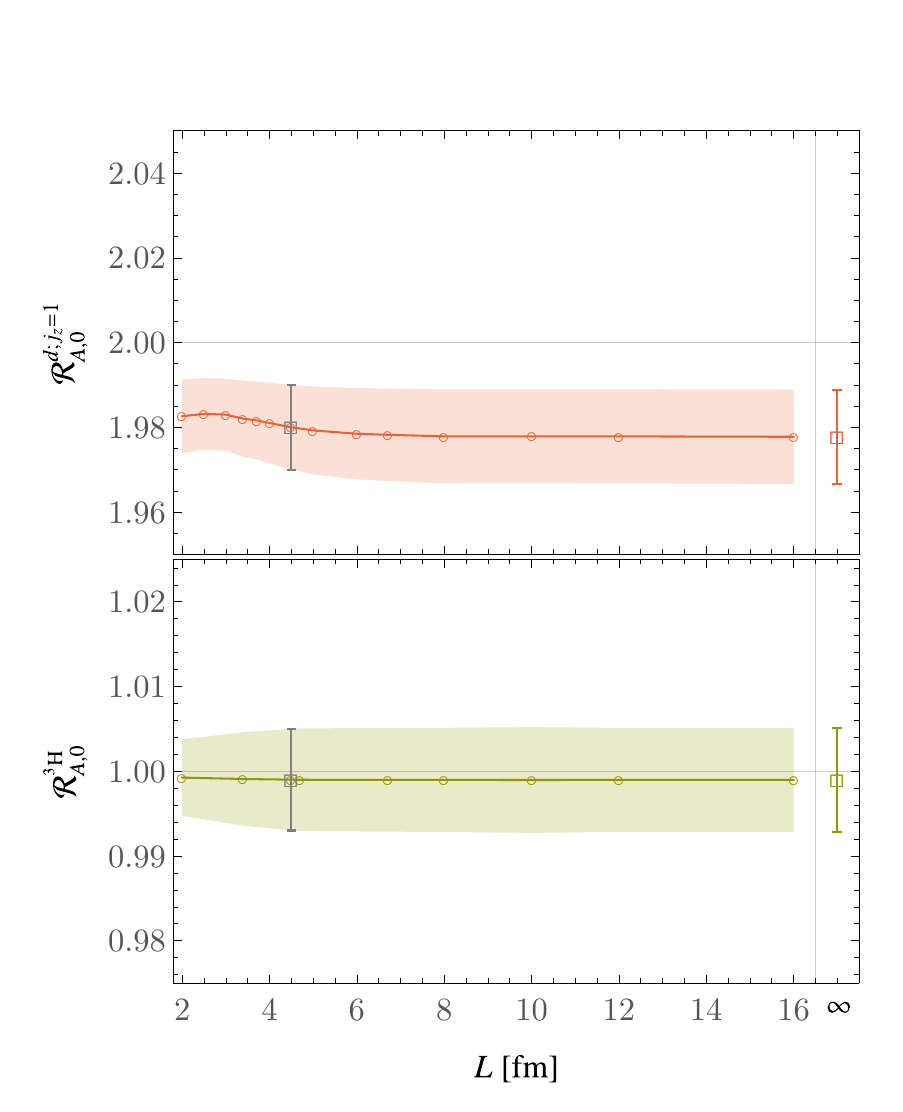}}
	\caption{\label{fig:axialisoscalar} (a) The dependence of the matrix elements of the isoscalar axial current in the  $j_z=1$ spin component of deuteron (upper) and $j_z=1/2$ component of  $\hthree$ (lower) on the coupling ratio $\widetilde{L}_{2,A}$.
	(b) The dependence of the matrix elements on the spatial extent of the lattice, $L$.  The details of the curves and points in the figure are as in Fig \ref{fig:axialisovec}.}
\end{figure*}

In order to evaluate the isovector axial current matrix elements, the EFT current in Eq.~\eqref{eq:axial_current} is used.
With the proton axial matrix element determining $g_A$ (up to exponentially small volume effects), ratios of matrix elements of the relevant current in both two and three-body states to that of the proton can be used to determine the two-body coupling ratio $\widetilde{L}_{1,A}=L_{1,A}/g_A$: 
\begin{align}
   {\cal R}_{A,3}^{np\leftarrow d}\equiv{}& \frac{2}{g_A} \frac{\langle \Psi_{np(1S0)} | {\cal A}_{3,3} | \Psi_{d;  j_z=0}\rangle}{\sqrt{\langle \Psi_{np(1S0)} | \Psi_{np(1S0)}\rangle \langle \Psi_{d; j_z=0} | \Psi_{d; j_z=0}\rangle}}
    \nonumber \\
   ={}& 2\left( 1 + \frac{\widetilde{L}_{1,A}}{2}  h_{np\leftarrow d}(\Lambda,L)  \right),
    \label{eq:pptoddude}
\end{align}
\begin{equation}
   {\cal R}_{A,3}^{\hthree}\equiv \frac{2}{g_A} \frac{\langle \Psi_{\hthree} | {\cal A}_{3,3} | \Psi_{\hthree}\rangle}{\langle \Psi_{\hthree} | \Psi_{\hthree}\rangle}
    =  \left( 1 + \frac{\widetilde{L}_{1,A}}{3}  h_{\hthree}(\Lambda,L) \right),
    \label{eq:gAhe3dude}
\end{equation}
where the spin-flavour structure of the states used to arrive at these expressions are given in Sec.~\ref{sec:svm}.

Figure~\ref{fig:axisovecvsL1A} shows the constraints that the LQCD calculations~\cite{Savage:2016kon} of the finite-volume matrix elements in the two channels place on the coupling combination $\widetilde{L}_{1,A}$. The consistency between the constraints in the two channels suggests that higher-order terms in the axial current (two-body operators with derivative insertions~\cite{Butler:2001jj}, or three-body operators) are suppressed as their power-counting would suggest. Note that this consistency is regulator-dependent. Were this to persist for physical quark masses, it would provide support for approaches to $pp$-fusion cross-section calculations that use tritium $\beta$-decay to constrain the relevant two-body LEC.

The values of $\widetilde{L}_{1,A}$ determined from each channel are scheme-dependent quantities, but can be combined with infinite-volume SVM wavefunctions to determine the infinite-volume matrix elements. Figure~\ref{fig:axisovecvsL} shows the infinite-volume extrapolation for both channels, and the extrapolated values are given in Table~\ref{tab:all} below.

Analogous analysis of the isoscalar axial matrix elements in the deuteron and $\hethree$ states allows for the determination of the two-body counterterm in Eq.~\eqref{eq:isoscalar_axial_current}. Ratios of the isoscalar axial current matrix element in the deuteron and $\hethree$ states to that in the proton state can be expressed as
\begin{align}
    {\cal R}_{A,0}^{d; j_z=1}\equiv{}&-\frac{2}{g_{A,0}} \frac{\langle \Psi_{d; j_z=1} | {\cal A}_{3,0} | \Psi_{d; j_z=1}\rangle}{\langle \Psi_{d; j_z=1} | \Psi_{d; j_z=1}\rangle} 
    \nonumber \\
    ={}& 2 \left( 1 - \widetilde{L}_{2,A} h_d(\Lambda,L)\right),
    \label{eq:disoscalardude}
\\[10pt]
    {\cal R}_{A,0}^{\hthree}\equiv{}&
    -\frac{2}{g_{A,0}} \frac{\langle \Psi_{\hthree } | {\cal A}_{3,0} | \Psi_{\hthree }\rangle}{\langle \Psi_{\hthree } | \Psi_{\hthree }\rangle} 
    \nonumber \\
    ={}& \left( 1 -\frac{2}{3} \widetilde{L}_{2,A} h_{\hethree}(\Lambda,L)\right).
    \label{eq:he3isoscalardude}
\end{align}
Figure~\ref{fig:axialisoscalar} shows the constraints on $\widetilde{L}_{2,A}$ obtained by matching to the LQCD calculation of Ref.~\cite{Chang:2017eiq}, and the corresponding infinite-volume extrapolation of the LQCD matrix elements. The extrapolated values are also reported in Table \ref{tab:all}. A mild tension is found between the values of $\widetilde{L}_{2,A}$ extracted from each matrix element, indicating the potential need for higher-order terms in the EFT description. 

\subsection{Magnetic moments}

\begin{figure*}[!t]
\includegraphics[width=0.9\columnwidth]{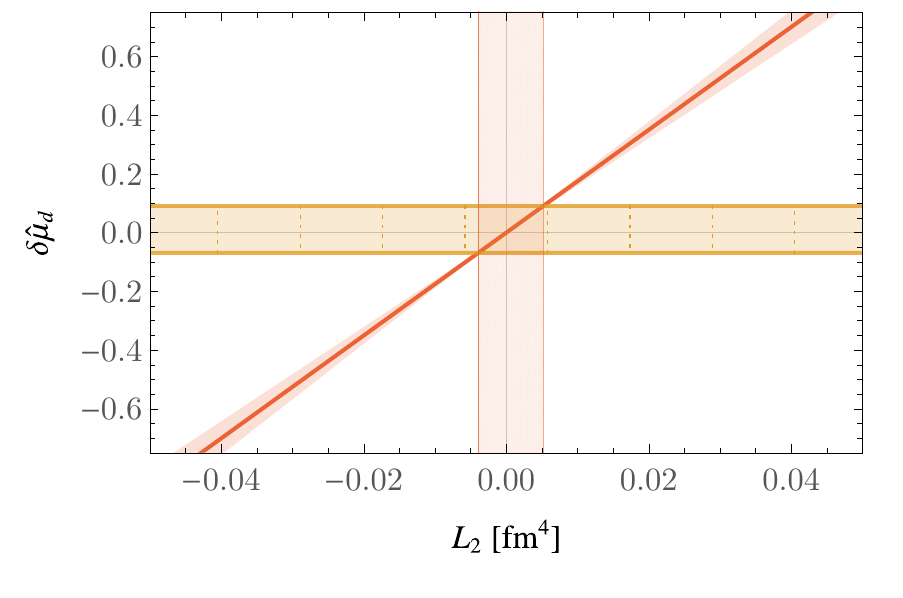}\quad
\includegraphics[width=0.9\columnwidth]{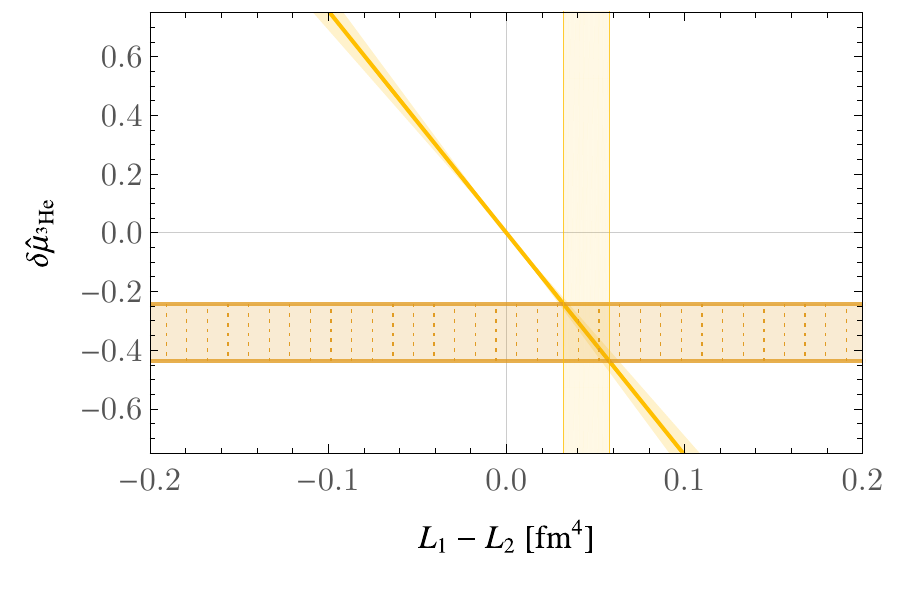}
\includegraphics[width=0.9\columnwidth]{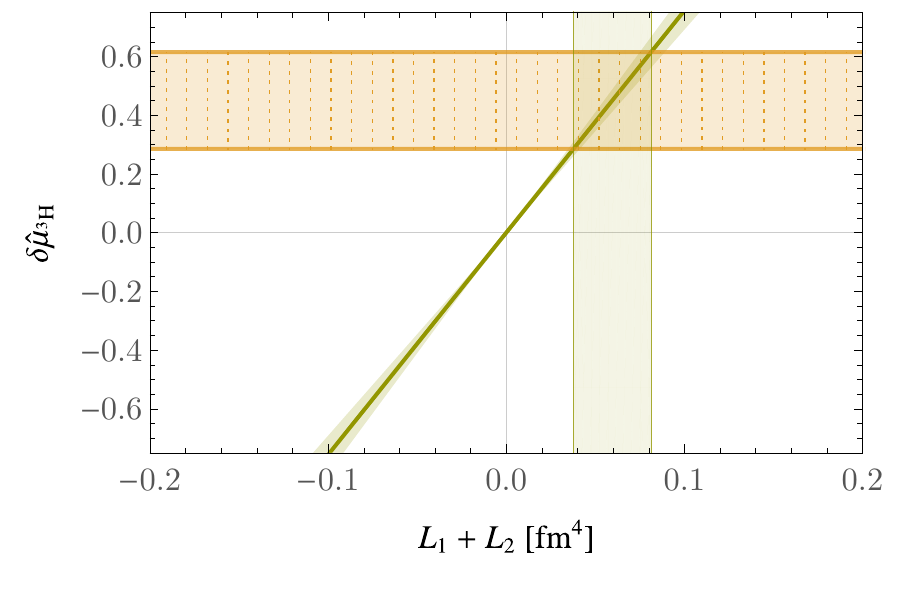}\quad
\includegraphics[width=0.9\columnwidth]{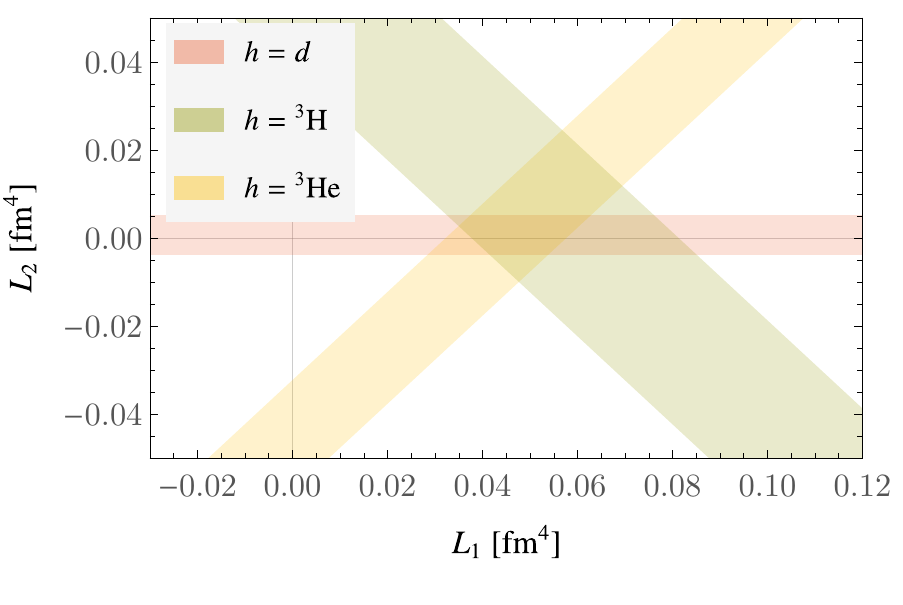}
	\caption{\label{fig:magmoments} The dependence of the various magnetic moment differences on the appropriate combinations of the two-body counterterms $L_{1,2}$ is shown in the upper row and the lower left panel. The lower right panel shows the combined constraints implied by agreement with the LQCD results of Ref.~\cite{Beane:2014ora}. The details of the curves and points in the figure are as in Fig \ref{fig:axialisovec}.}
\end{figure*}

The magnetic moments of light nuclei have been extracted from the linear response of LQCD calculations to a constant background magnetic field oriented in the $z$-direction \cite{Beane:2014ora,Beane:2015yha}. 
In EFT, these quantities are determined by the couplings in Eq.~\eqref{eq:JEM}. The differences between the magnetic moments of the deuteron, ${}^3$H, and ${}^3$He states and the relevant naive shell-model predictions in terms of proton and neutron magnetic moments can be expressed as:
\begin{align}
    \delta \hat\mu_d \equiv{} & \hat\mu_d -(\hat\mu_p+\hat\mu_n)
    \nonumber \\
    ={}& \frac{2M_N}{e}\frac{\langle \Psi_{d; j_z=1} | {\cal J}^{EM}_3 | \Psi_{d; j_z=1}\rangle}{\langle \Psi_{d; j_z=1} | \Psi_{d; j_z=1}\rangle} -2\kappa_0 \nonumber \\
    ={}&
    2M_N L_2 h_d(\Lambda,L), 
    \\[10pt]
    \delta \hat\mu_{\hthree} \equiv{} & \hat\mu_{\hthree} -\hat\mu_n
    = \frac{2M_N}{e}\frac{\langle \Psi_{\hthree} | {\cal J}^{EM}_3 | \Psi_{\hthree}\rangle}{\langle \Psi_{\hthree} | \Psi_{\hthree}\rangle} -\kappa_n \nonumber \\
    ={}&
    \frac{M_N}{3} (L_1 + L_2)h_{\hthree}(\Lambda,L), 
    \\[10pt]
    \delta \hat\mu_{\hethree} \equiv{} & \hat\mu_{\hethree} -\hat\mu_p
    = \frac{2M_N}{e}\frac{\langle \Psi_{\hethree} | {\cal J}^{EM}_3 | \Psi_{\hethree}\rangle}{\langle \Psi_{\hethree} | \Psi_{\hethree}\rangle} -\kappa_p \nonumber \\
    ={}&
   -\frac{M_N}{3} (L_1 - L_2)h_{\hethree}(\Lambda,L), 
\end{align}
where $\hat\mu_h$ is the magnetic moment of hadron $h$ in natural nuclear magnetons $e/2M_N$ defined using the nucleon mass at the quark masses of the lattice calculations, $M_N=1.634(18)$~GeV~\cite{Beane:2012vq}. 

\begin{figure}[!t]
\includegraphics[width=0.9\columnwidth]{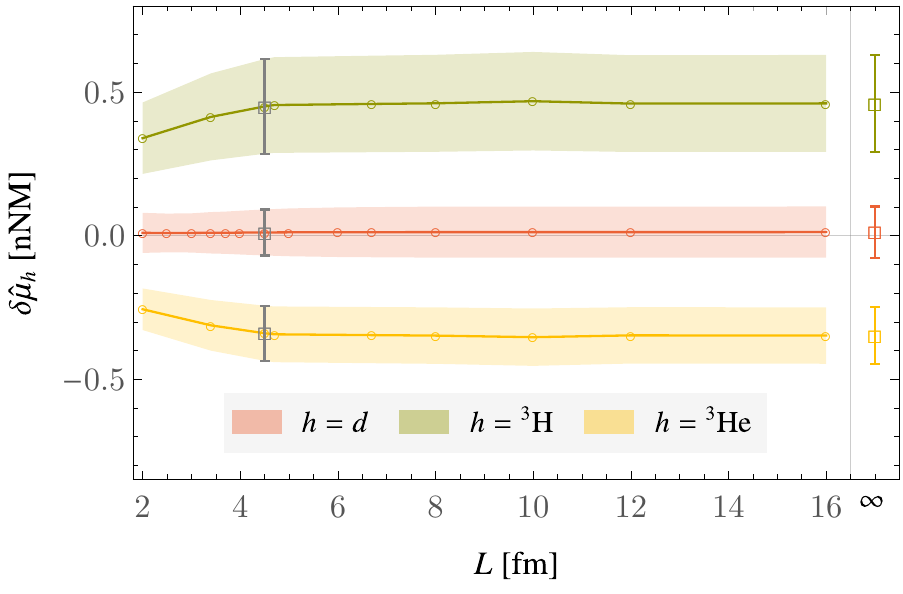}
	\caption{\label{fig:magmomentsLextrap} The infinite-volume extrapolations of the magnetic moment differences for $d$, $\hthree$ and $\hethree$.}
\end{figure}

In Fig.~\ref{fig:magmoments}, the LQCD calculations of $\delta \hat\mu_h$ for $h\in\{d,\hthree,\hethree\}$ in Ref.~\cite{Beane:2014ora} are used to constrain the EFT couplings $L_{1,2}$.\footnote{Note that in Ref.~\cite{Beane:2014ora} the magnetic background field does not couple to sea quarks, so only isovector quantities are calculated completely; the error from this quenching of the magnetic field is ignored here. In principle, the isovector $np\to d \gamma$ M1 transition can also be used to constrain $L_1$, but it is not used in this work.} Since the magnetic moment differences are dependent of various combinations of the couplings, the constraints take the form of bands in the $L_1$--$L_2$ plane as shown in the figure. All three constraints are seen to be consistent for a range of values of the couplings. This determination of the LECs allows for extrapolation of the magnetic moments to infinite volume as shown in Fig.~\ref{fig:magmomentsLextrap}, and produces the values shown in Table~\ref{tab:all}. 

\subsection{Scalar matrix elements: nuclear $\sigma$ terms}

Ratios of the matrix elements of the isoscalar and isovector scalar currents in hadron $h$ to those in the proton can be expressed as
\begin{align}
    {\cal R}_{S,0}^{h}\equiv&
    \frac{1}{g_{S,0}}\frac{\langle \Psi_{h} | {\cal S}_{0} | \Psi_{h}\rangle}{\langle \Psi_{h} | \Psi_{h}\rangle}
= \left( A_h - \frac{f_{S,0}^h}{2g_{S,0}}h_h(\Lambda,L) \right), 
    \label{eq:scalar_isoscalar_dude} 
\intertext{and}
   {\cal R}_{S,3}^{h}\equiv&
   \frac{1}{g_{S,3}}\frac{\langle \Psi_{h} | {\cal S}_{3} | \Psi_{h}\rangle}{\langle \Psi_{h} | \Psi_{h}\rangle}
= \left(2T_{3}^h - \frac{f_{S,3}^h}{g_{S,3}}h_h(\Lambda,L) \right), 
    \label{eq:scalar_isovector_dude}
\end{align}
where $A_h$ denotes the atomic number of the nucleus, $T_3^h$ is its third component of isospin,  and
\begin{align}
    f_{S,0}^h = \begin{cases}
    \widetilde{C}_0+\widetilde{C}_1, & h=d \\
    \widetilde{C}_0-3\widetilde{C}_1, & h=pp \\
    \widetilde{C}_0-\widetilde{C}_1, & h={}^3\text{H}
    \end{cases},
    \qquad  
    f_{S,3}^h =\begin{cases}
   \widetilde{C}_V
    , & h=pp \\
    \widetilde{C}_V, & h={}^3\text{H}
    \end{cases}.
\end{align}

\begin{figure*}[!p]
	\subfigure[]{\includegraphics[width=0.85\columnwidth]{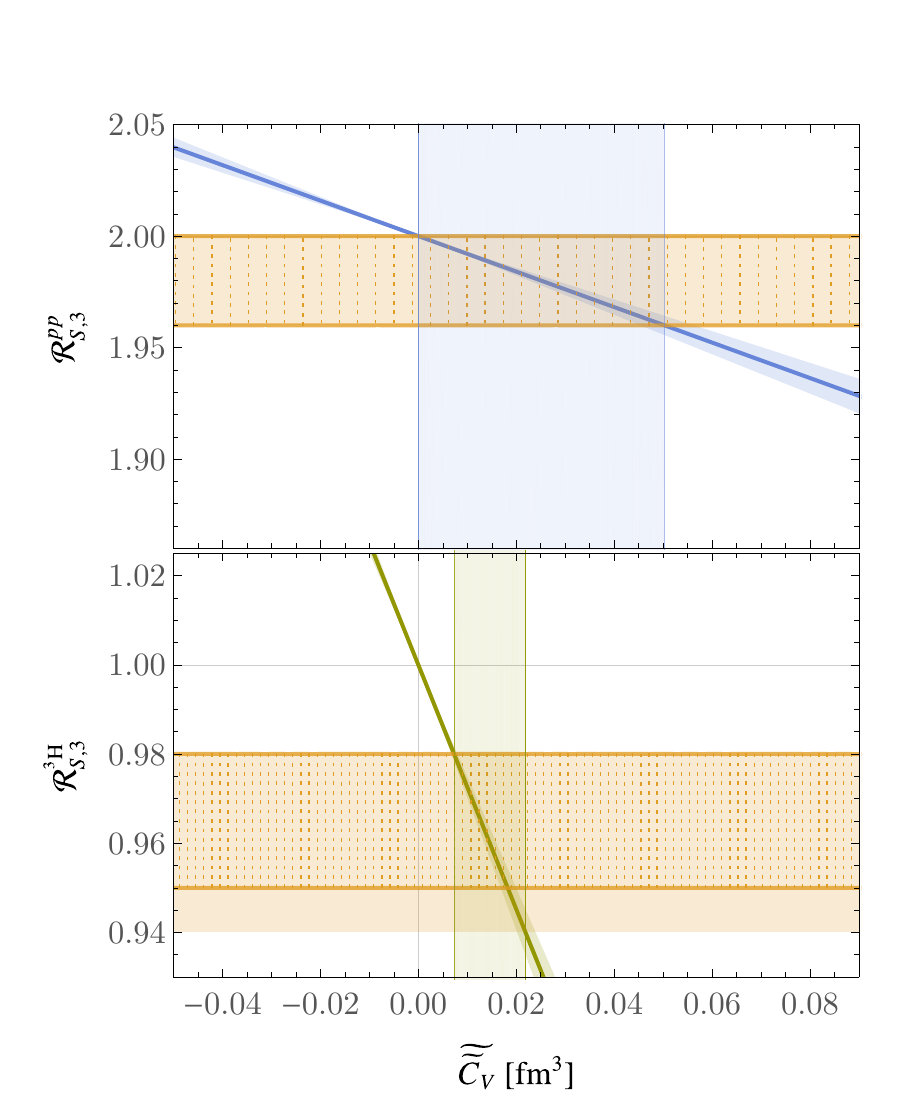}}
	\subfigure[]{\includegraphics[width=0.85\columnwidth]{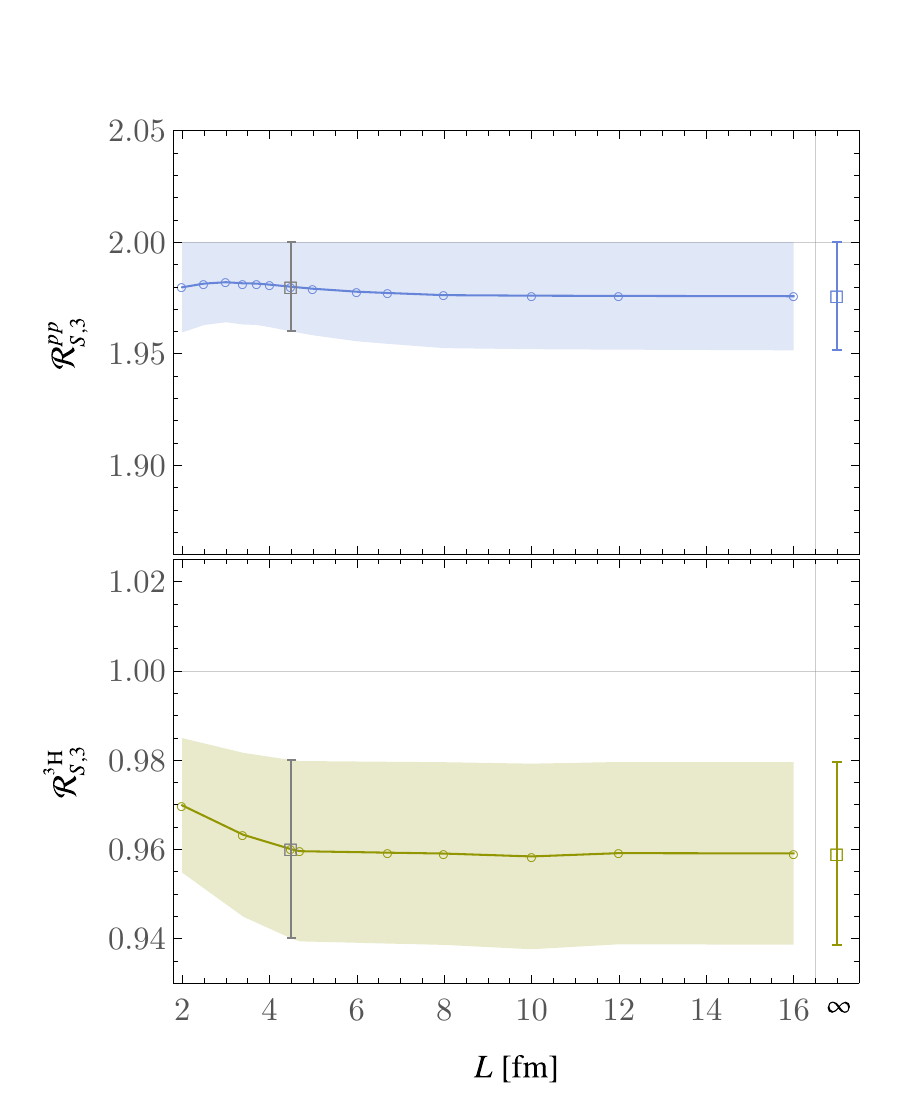}}
	\caption{\label{fig:isovectorscalar} (a) The dependence of the ratios of the $pp$ (upper) and $\hthree$ (lower) isovector scalar matrix elements to that of the proton on the LEC ratio $\widetilde{\widetilde{C}}_{V}$. 
	(b) The infinite-volume extrapolation of these ratios after constraining the LEC ratio. The details of the curves and points in the figure are as in Fig~\ref{fig:axialisovec}.}
\vspace{5mm}
\includegraphics[width=0.85\columnwidth]{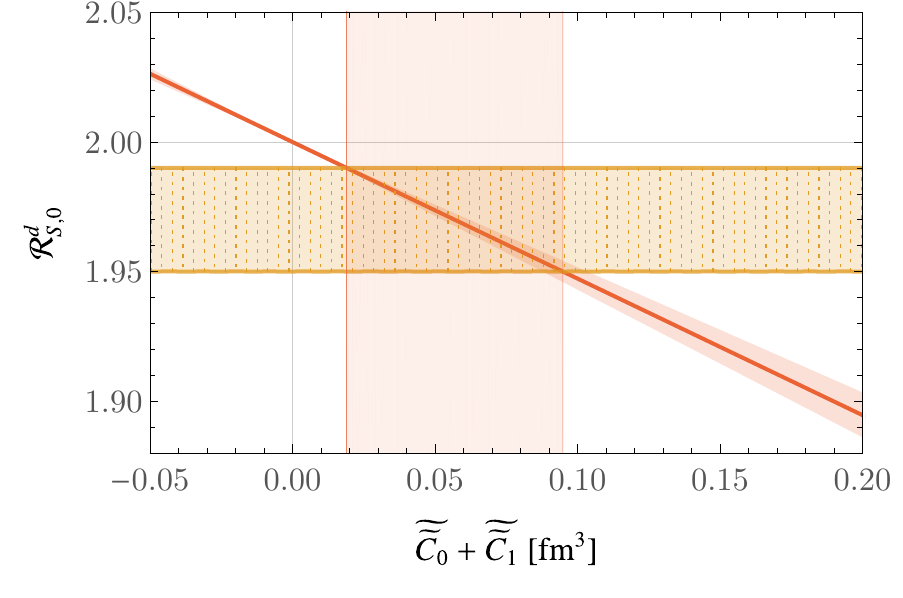}\quad
\includegraphics[width=0.85\columnwidth]{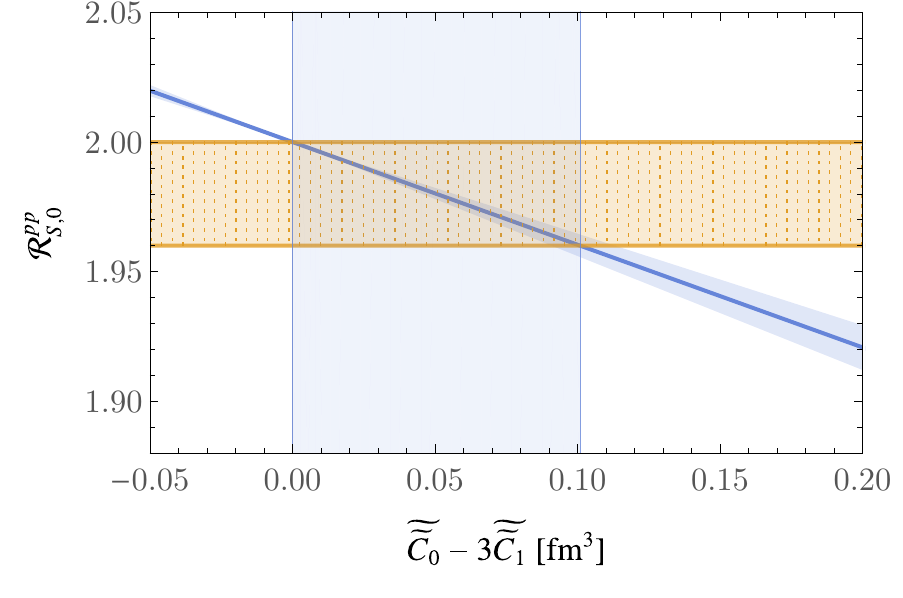}
\includegraphics[width=0.85\columnwidth]{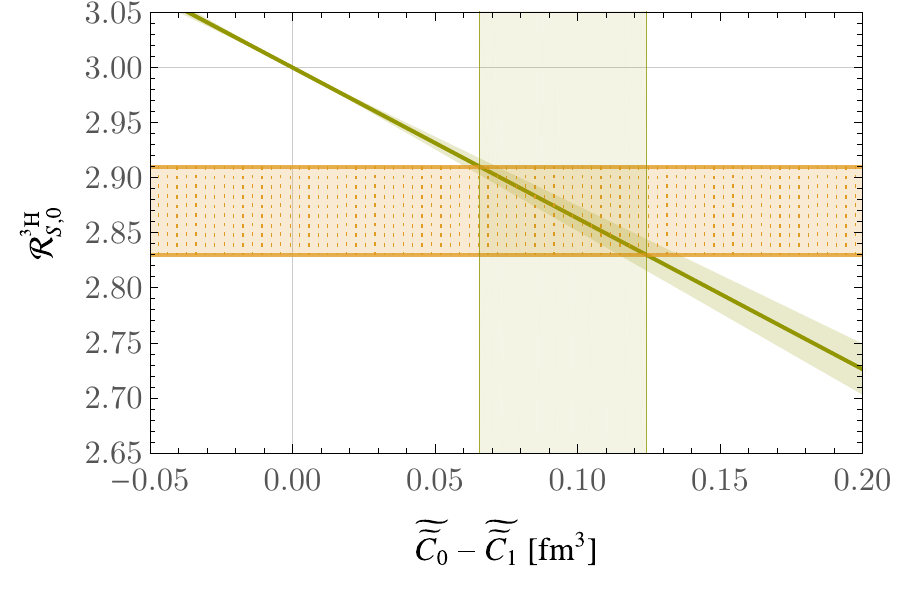}\quad
\includegraphics[width=0.85\columnwidth]{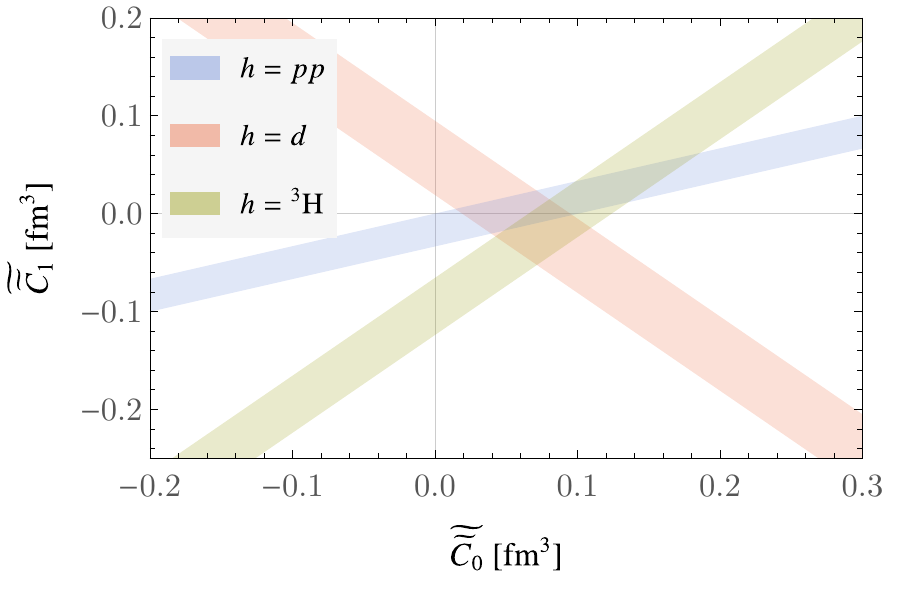}
	\caption{\label{fig:isoscalarscalar} The isoscalar scalar current matrix element ratios for $d$, $pp$ and $\hthree$ as a function of the relevant combinations of LEC ratios $\widetilde{\widetilde{C}}_{0,1}$.
	The lower right panel shows the resulting constraints on these LEC ratios. The details of the curves and points in the figure are as in Fig \ref{fig:axialisovec}. }
\end{figure*}

The quantity $\widetilde{\widetilde{C}}_{V}=\widetilde{C}_{V}/g_{S,3}$ is constrained by comparing Eq.~\eqref{eq:scalar_isovector_dude} to LQCD calculations of ratios of isovector scalar current matrix elements in different nuclei from Ref.~\cite{Chang:2017eiq}. The results of this comparison are shown in Fig.~\ref{fig:isovectorscalar}. Determinations of $\widetilde{\widetilde{C}}_{V}$ from both the $pp$ and $\hthree$ systems are consistent, with $\hthree$ providing a considerably more stringent constraint. The extracted values of the LEC are also used to extrapolate the LQCD matrix elements to infinite volume, as shown in the figure and presented in Table~\ref{tab:all}.

\begin{figure}[!t]
\includegraphics[width=0.9\columnwidth]{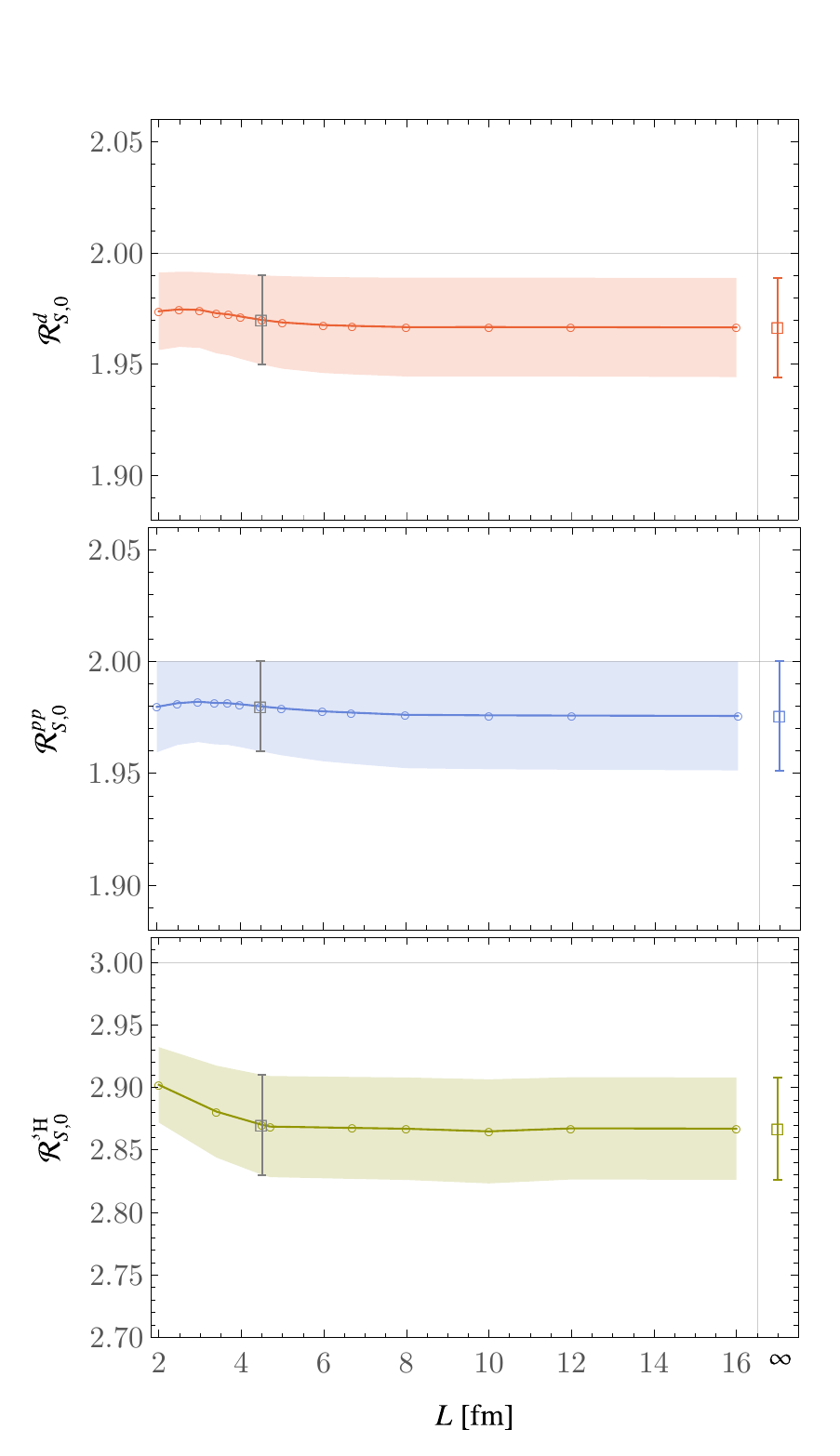}
	\caption{\label{fig:isoscalarscalarLextrap} The infinite-volume extrapolations of the isoscalar scalar current matrix element ratios for $d$, $pp$ and $\hthree$.}
\end{figure}

Similarly Fig.~\ref{fig:isoscalarscalar} compares LQCD calculations of the ratio of the scalar isoscalar current matrix element in nuclei to that in the proton to the expectations of Eq.~\eqref{eq:scalar_isoscalar_dude} for the deuteron, diproton and $\hthree$. The three states provide sufficient information to constrain the LEC ratios $\widetilde{\widetilde{C}}_{0,1}= \widetilde{C}_{0,1}/g_{S,0}$ and the constrained values are used to extrapolate the LQCD matrix elements to infinite volume as shown in Fig.~\ref{fig:isoscalarscalarLextrap} and presented in Table \ref{tab:all}. As noted in Sec.~\ref{subsec:scalarEFT}, the couplings $\widetilde{C}_{0,1}$ that occur in the isoscalar scalar current are related to the Lagrangian counterterms $C_{0,1}$ in the limit of massless quarks.

\subsection{Tensor  matrix elements}

For the tensor current, isoscalar matrix element ratios to those in the nucleon are given by
\begin{align}
    {\cal R}_{T,0}^{h}\equiv&\frac{2}{g_{T,0}}\frac{\langle \Psi_{h} | {\cal T}_{12,0} | \Psi_{h}\rangle}{\langle \Psi_{h} | \Psi_{h}\rangle}
    =  \left( 2S_3^h -  f^h_{T,0}h_h(\Lambda,L) \right),
    \label{eq:tensor_isoscalar_dude}
\end{align}
for $h\in\{d,\hthree\}$,
where $S_3^h$ is the third component of spin and 
\begin{equation}
   f_{T,0}^h =\begin{cases}
   \widetilde{L}_{2,T}
    , & h=d; j_z=1 \\
    \frac{1}{3}\widetilde{L}_{2,T}, & h=\hthree
    \end{cases},
\end{equation}
where $\widetilde{L}_{2,T}=L_{2,T}/g_{T,0}$.

For the isovector case, the corresponding ratio in $\hthree$ is
\begin{align}
 {\cal R}_{T,3}^{\hthree}\equiv&
     \frac{2}{g_{T,3}}\frac{\langle \Psi_{{}^3\text{H}} | {\cal T}_{12,3} | \Psi_{{}^3\text{H}}\rangle}{\langle \Psi_{{}^3\text{H}} | \Psi_{{}^3\text{H}}\rangle}
= \left( 1 +  \frac{\widetilde{L}_{1,T}}{3}h_{{}^3\text{H}}(\Lambda,L) \right),
    \label{eq:tensor_isovector_dude}
\end{align}
in terms of the LEC ratio $\widetilde{L}_{1,T}=L_{1,T}/g_{T,3}$

\begin{figure*}[!t]
	\subfigure[]{\includegraphics[width=0.9\columnwidth]{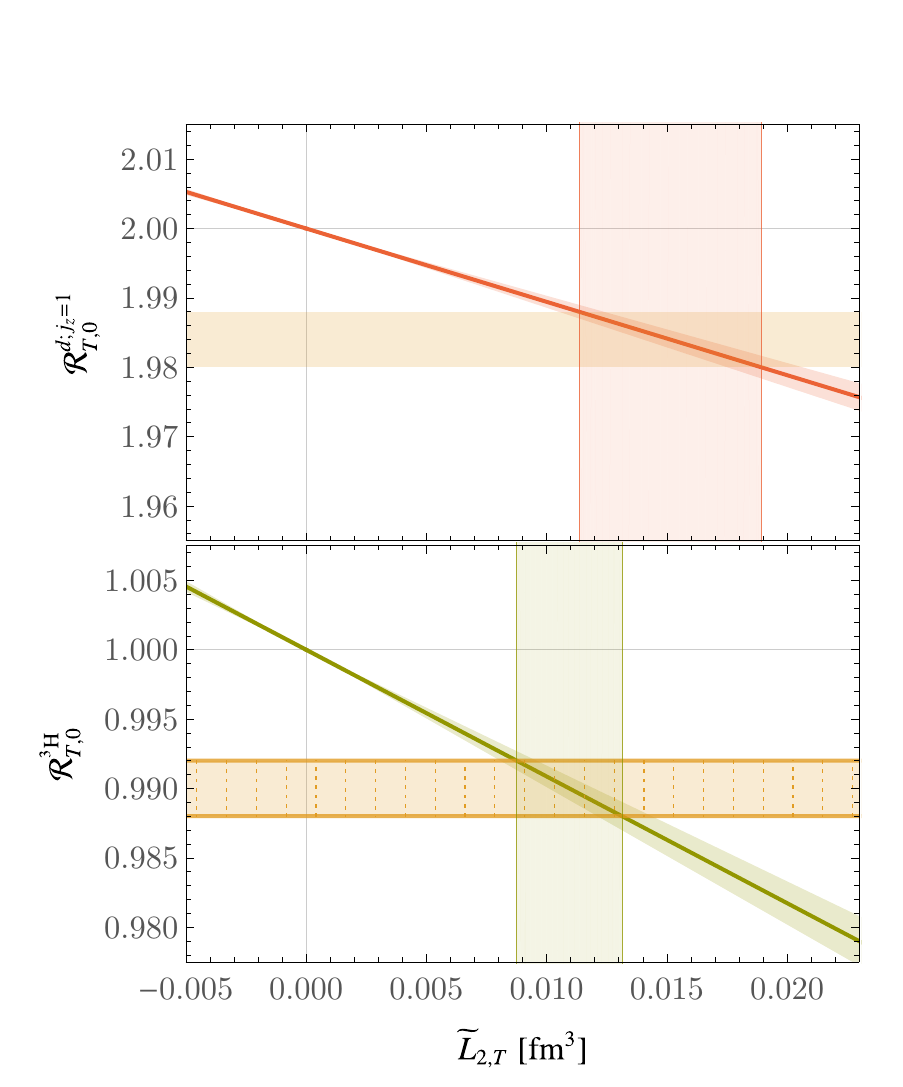}}
	\subfigure[]{\includegraphics[width=0.9\columnwidth]{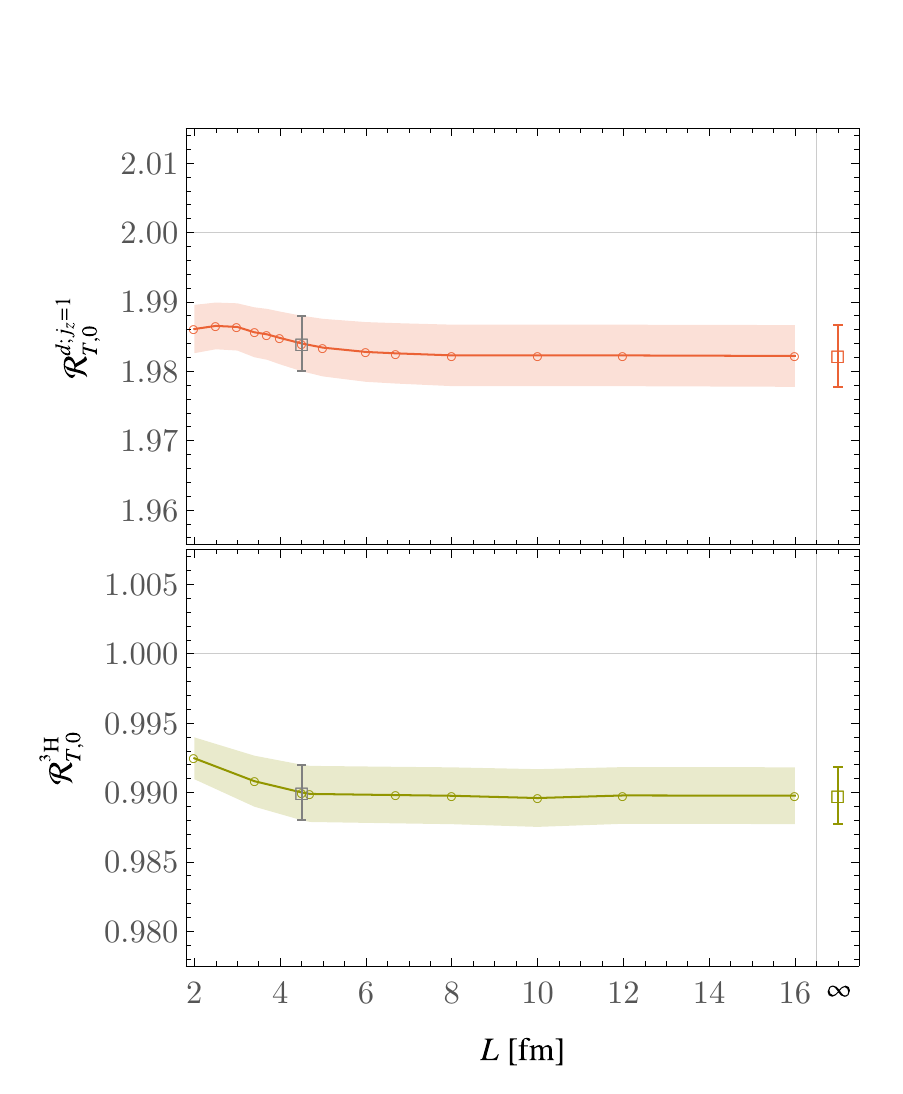}}
	\caption{\label{fig:isoscalartensorvsL} (a) The dependence of the $d$ (upper) and $ \hthree$ (lower) isoscalar tensor  matrix elements on $\widetilde{L}_{2,T}$. 
	(b) The dependence of these matrix elements on the lattice extent, $L$. The details of the curves and points in the figure are as in Fig \ref{fig:axialisovec}.}
\end{figure*}

\begin{figure*}[!t]
		\subfigure[]{\includegraphics[width=0.9\columnwidth]{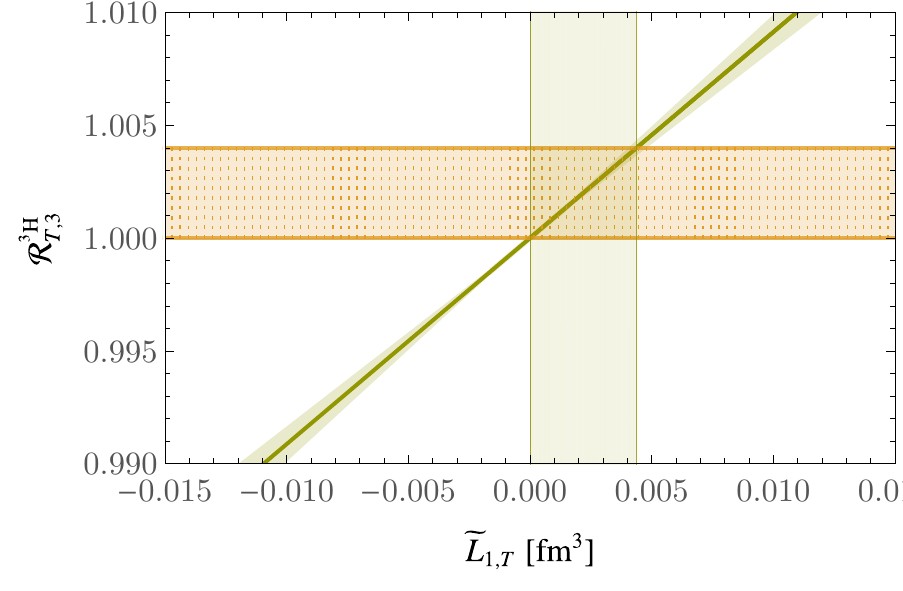}}
	\subfigure[]{\includegraphics[width=0.9\columnwidth]{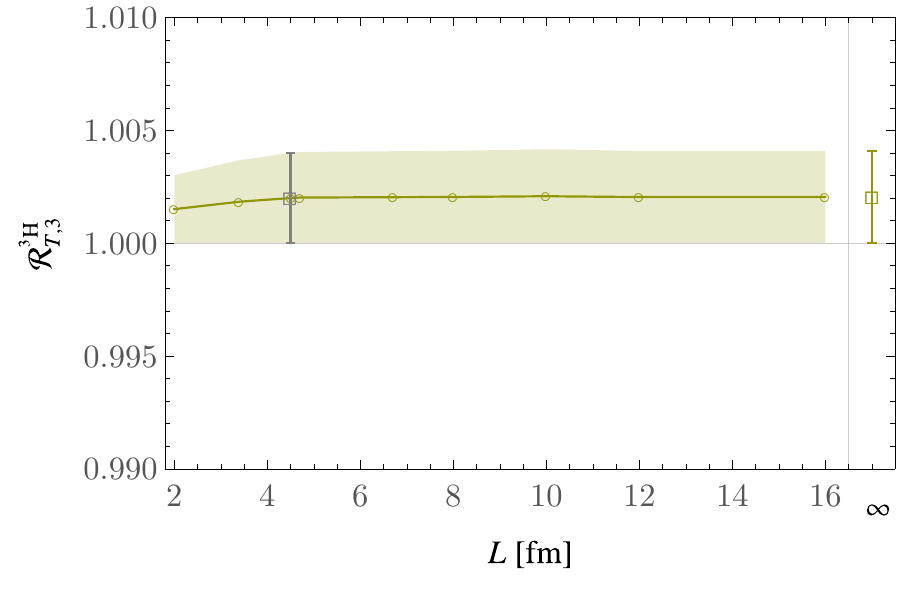}}
	\caption{\label{fig:isovectortensorvsL} (a) The dependence of the $\hthree$ isovector tensor transition matrix element on $\widetilde{L}_{1,T}$. 
	(b) The dependence of this matrix element on the lattice extent, $L$. The details of the curves and points in the figure are as in Fig \ref{fig:axialisovec}. }
\end{figure*}

Figures \ref{fig:isoscalartensorvsL} and \ref{fig:isovectortensorvsL} show the comparisons of Eqs.~\eqref{eq:tensor_isoscalar_dude} and \eqref{eq:tensor_isovector_dude} to the respective LQCD calculations \cite{Chang:2017eiq}. In the isoscalar case, consistency is seen in the values of $\widetilde{L}_{2,T}$ that arise from comparison to either $d$ or $\hthree$ matrix element ratios from LQCD. As for the  matrix elements of other operators considered above, the constrained LECs enable infinite volume extrapolations of the matrix elements which are presented in Table~\ref{tab:all}.

\subsection{Twist-two operators: the quark momentum fraction}

The first moment of the isovector unpolarised parton distribution has been computed in LQCD for nuclei with $A\in\{2,3\}$ in Ref.~\cite{Detmold:2020snb}, and correspond to the difference in the longitudinal momentum fractions carried by up and down quarks. In the finite-volume SVM, the matrix elements of the operators in Eq.~\eqref{eq:twist2ops} contain no spin structure and so the finite-volume matching and extrapolation is analogous to that on the isovector scalar current above. In particular, ratios of the isovector matrix element in hadron $h$ to the proton matrix elements are given by
\begin{align}
{\cal R}_{{\cal O}^n,3}^{h}\equiv&  \frac{A^h}{(Z^h-N^h)\langle  x^n\rangle_3}\frac{\langle \Psi_{h} | {\cal O}^{(n)}_{3} | \Psi_{h}\rangle}{\langle \Psi_{h} | \Psi_{h}\rangle}
\nonumber \\
    =&  \left(1 + \frac{\alpha_{n,3}}{(Z^h-N^h)\langle x^n\rangle_{3}}h_h(\Lambda,L) \right),
    \label{eq:scalar_isovector_dude}
\end{align}
for $h\in\{pp,\hthree\}$.

\begin{figure*}[!t]
	\subfigure[]{\includegraphics[width=0.9\columnwidth]{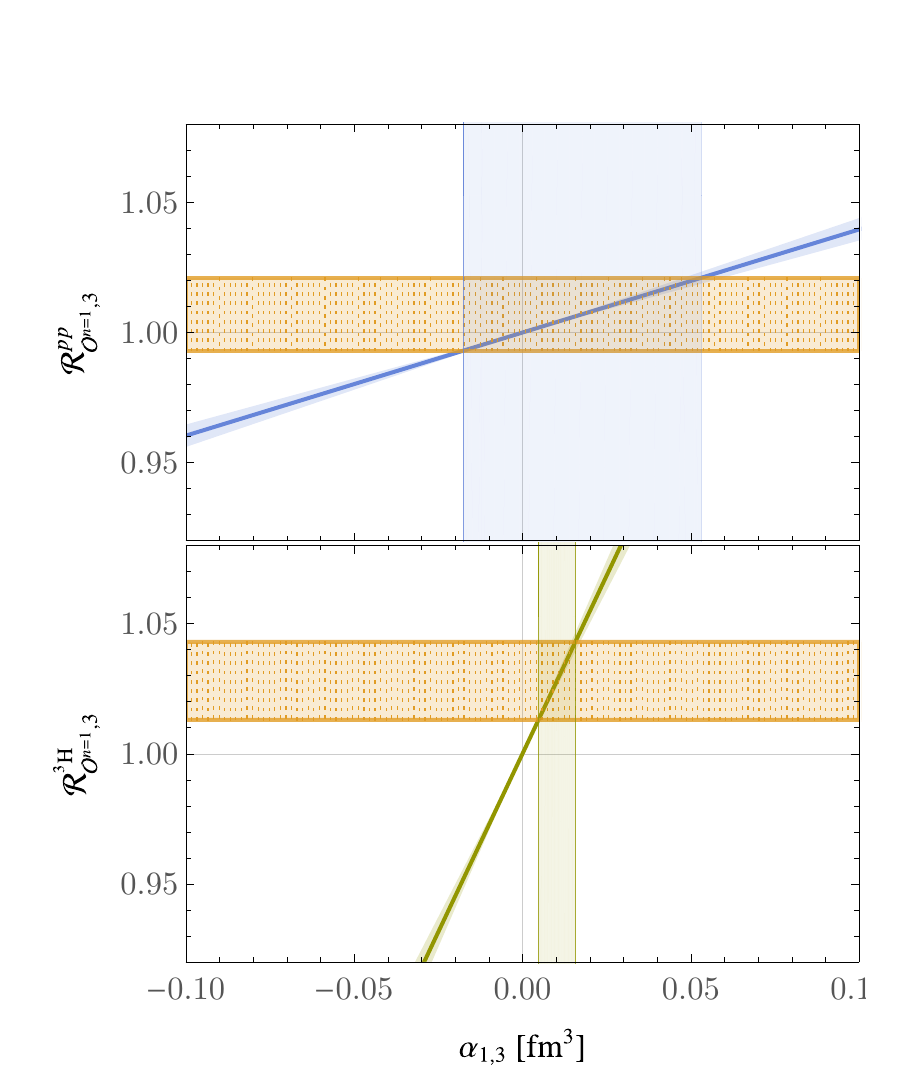}}
	\subfigure[]{\includegraphics[width=0.9\columnwidth]{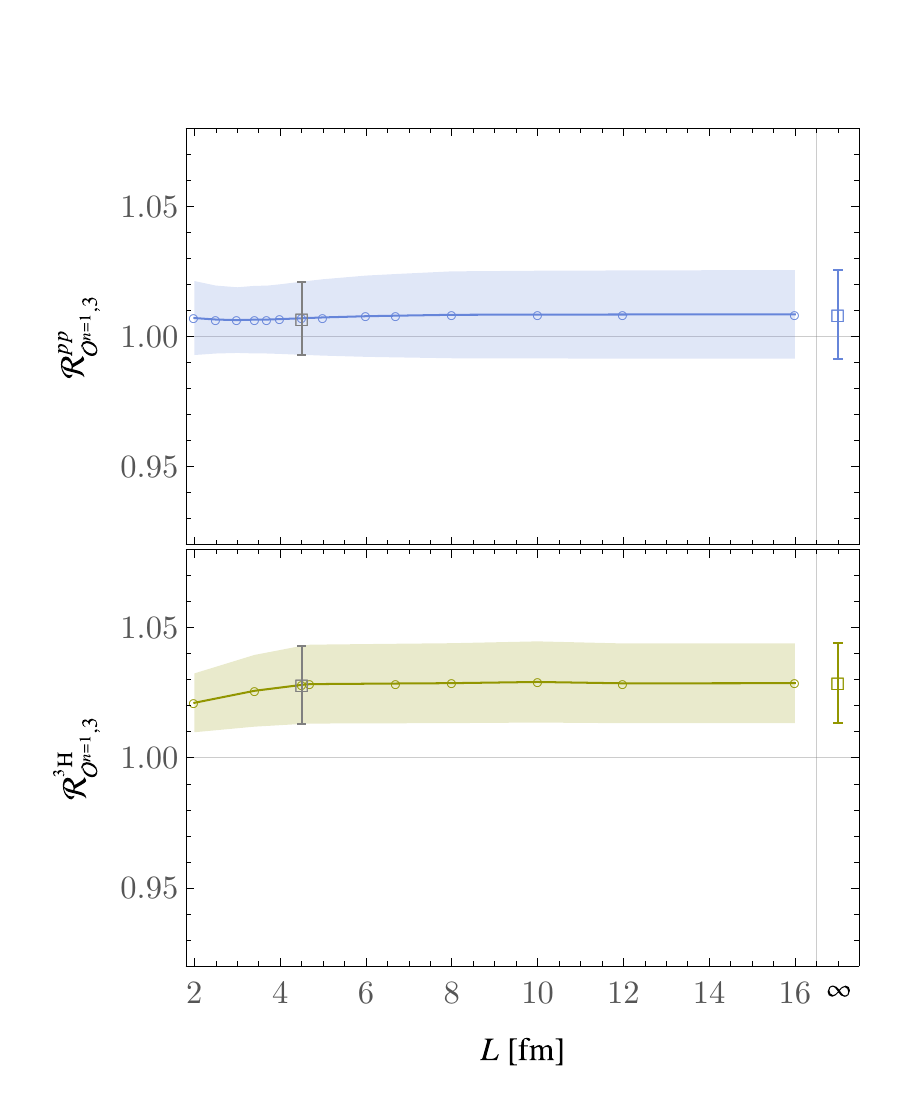}}
	\caption{\label{fig:isovectortwist2} (a) The dependence of the  isovector twist-two matrix elements in $pp$ and $\hthree$ on the two-body LEC, $\alpha_{1,3}$.
	(b) The dependence of the matrix element on the lattice extent, $L$.  The details of the curves and points in the figure are as in Fig \ref{fig:axialisovec}. }
\end{figure*}

LQCD data in a calculation with $L=4.5$~fm constrain the single-nucleon isovector momentum fraction and two-nucleon counterterm $\alpha_{1,3}$, as shown in Fig.~\ref{fig:isovectortwist2}. The extrapolated  infinite-volume matrix elements are reported in Table \ref{tab:all}.

\begin{table}[!t]
    \centering
    \begin{ruledtabular}
\begin{tabular}{cccc}
 Quantity ${\cal O}$ &State $h(\to h^\prime)$  & ${\cal O}(L=4.5\ {\rm fm})$ & ${\cal O}(L=\infty)$ \\
\hline 

{$\delta\hat\mu^{(h)}$}&&  \cite{Beane:2014ora}  \\
\hline
&$d\left(j_{z}=\pm 1\right)$ &   0.011(80) & 0.012(89)\\
&${}^3 \mathrm{He}$ & -0.34(10) & -0.35(10)\\
&${}^3 \mathrm{H}$ &  0.45(16) & 0.46(17)\\
\hline
{${\cal R}_{A,3}^h$} && \cite{Savage:2016kon}\\ \hline
&$\hthree$ &   0.979(10) & 0.978(11)\\ \hline
&$np(\to d)$ &  1.978(31) & 1.975(36) \\ \hline
{${\cal R}_{A,0}^h$ } && \cite{Chang:2017eiq}\\ \hline
&$d$ & $1.98(1)$ & 1.98(1) \\ 
&$\hthree$ & $0.999(6)$ & 0.999(6)\\ \hline
{${\cal R}_{T,3}^h$} && \cite{Chang:2017eiq}\\ \hline
&$\hthree$ & $1.002(2)$ & 1.002(2)\\ \hline
& $np(\to d)$ & -- & -1.415(2) \\ \hline
{${\cal R}_{T,0}^h$ } && \cite{Chang:2017eiq} \\ \hline
&$d$ & $1.984(4)$ & 1.982(4)\\ 
&$\hthree$ & $0.990(2)$ & 0.990(2)\\ \hline
{${\cal R}_{S,3}^h$} && \cite{Chang:2017eiq}\\ \hline
&$pp$ & $1.98(2)$ & 1.98(2) \\ 
&$\hthree$ & $0.96(2)$ & 0.96(2) \\ \hline
{${\cal R}_{S,0}^h$ } && \cite{Chang:2017eiq}\\ \hline
&$pp$ & $1.98(2)$ & 1.98(2)\\ 
&$d$ & $1.97(2)$ & 1.97(2)\\ 
&$\hthree$ & $2.87(4)$ & 2.87(4)\\ \hline
{ ${\cal R}^h_{{\cal O}^{n=1},3}$ } && \cite{Detmold:2020snb}\\ \hline
&$pp$ & $1.007(14)$ & 1.008(17)\\ 
&$\hthree$ & $1.028(15)$ & 1.029(15)\\ 
\end{tabular}
\end{ruledtabular}
\caption{\label{tab:all}Nucleus-to-proton ratios of quantities computed at $m_\pi=806$~MeV in a $L=4.5$~fm volume and extrapolated to infinite volume. Ratios are computed from data presented in the reference shown at the top of each section of the table. Where multiple uncertainties are given in the literature, they have been combined in quadrature and standard error propagation has been employed in cases where the ratios are not given in the original works.
For ${\cal R}_{T,3}^{d\leftarrow np}$, the LEC determined from the $\hthree$ channel is used to make a prediction, as there are no LQCD results available.
}
\end{table}

\section{Discussion}
\label{sec:discuss}

Finite-volume pionless effective field theory implemented through the stochastic variational method has been used to extrapolate EFT wavefunctions and matrix elements for $A\in\{2,3\}$ nuclei, matched to LQCD calculations in a finite lattice volume, to infinite volume. This numerical approach can effectively describe bound-state systems, can cleanly reproduce scattering states, and furthermore exhibits volume scaling that is consistent with the predictions of the L\"uscher approach to high accuracy.
To some degree, the method circumvents the complexities of analytic approaches generalising that of L\"uscher for two-body systems to larger number of particles. However, as the atomic number of the system increases, the finite-volume SVM scales relatively poorly, and it does not seem practical to extend much beyond $A=4$ systems or more than three-body interactions. Given that the use of the finite-volume aspect of the method is tailored to match LQCD calculations, for which increasing $A$ is also costly, this is perhaps not a significant limitation: the finite-volume SVM can be used to determine EFT counterterms which can then be used in the infinite-volume SVM (or other many-body methods) to perform calculations for larger nuclei. 

For all of the matrix elements studied in this work, which include isoscalar and isovector scalar, axial, and tensor matrix elements, as well as magnetic moments and the isovector longitudinal momentum fraction, it is found that for the large quark masses used in the LQCD calculations, the lattice volume of $L=$4.5~fm as used for the calculations is large enough that there are essentially no finite-volume corrections. At lighter quark masses, however, one might anticipate that larger lattice volumes will be required to achieve the same behaviour. For almost all of the matrix elements investigated in this work, it is also notable that although the constraints from LQCD calculations of three-body systems are typically tighter, the EFT$_{\pislash}$ LECs determined from the LQCD calculations of two and three-body systems are consistent to within one standard deviation (with the notable exception of the isoscalar axial LEC $L_{2,A}$ in Eq.~\eqref{eq:isoscalar_axial_current}, for which there is a slight tension). This indicates that higher-order terms in the relevant currents are suppressed in the exponential regulator scheme used in the SVM at these values of the quark masses. In some cases, this suppression has particular consequence; for example, for the isovector axial matrix elements, the LEC $L_{1,A}$ determined from measurements of tritium $\beta$-decay is used in calculations of the $pp$-fusion cross-section \cite{De-Leon:2016wyu}. With the LECs determined from LQCD calculations, predictions can be made for other quantities for which there are no LQCD results; in Table \ref{tab:all}, as an example, the $np\leftarrow d$ tensor transition matrix element is predicted from the LEC determined from the triton matrix element of the same current.

Since the finite-volume SVM provides representations of low-lying excited states as well as the ground states that have been the focus of this work, it can also be used to match matrix elements of second-order current insertions such as in double-$\beta$ decay. In such processes, a sum over excited nuclear states occurs for times between those of the two currents, with the matrix elements of interest being
\begin{equation}
   \sum_n \frac{\langle \Psi_f | {\cal J} | \Psi_n\rangle\langle \Psi_n | {\cal J} | \Psi_i\rangle}{E-E_n}.
\end{equation}
In the finite-volume EFT context this corresponds to inclusion of the discrete states in principle up to energy-scale of the EFT cutoff. 
In order to accurately represent these contributions, care must be taken that these states are equivalently well optimised. This will require significant numerical effort, but is likely to be feasible. 
Ultimately, the finite-volume SVM appears to be a powerful tool to capitalise on LQCD calculations of systems with small $A$, which are approaching a novel era of systematic control.

\section*{Acknowledgments}

We are grateful to J-W. Chen, Z. Davoudi, M. Illa, A. Parre\~no, M. J. Savage, and M. Wagman for discussions. The authors are supported in part by the U.S.~Department of Energy, Office of Science, Office of Nuclear Physics under grant Contract Number DE-SC0011090 and by the Carl G and Shirley Sontheimer Research Fund. WD is also supported within the framework of the TMD Topical Collaboration of the U.S.~Department of Energy, Office of Science, Office of Nuclear Physics, and  by the SciDAC4 award DE-SC0018121. PES is additionally supported by the National Science Foundation under EAGER grant 2035015, by the U.S. DOE Early Career Award DE-SC0021006, and by a NEC research award.

\appendix

\section{Alternate form for currents}
\label{app:basis}

Currents are derived by first considering the relativistic form at the quark level, matching onto relativistic nucleon operators with the same C, P and T properties, and then performing a nonrelativistic reduction.
In the main text, the various currents were written using the projectors in Eq.~\eqref{eq:projector}. They can also be written in terms of Pauli matrices in spin and isospin as follows, where each expression is given only to the order used in this work:
\begin{align}
A_{i,a}={}& \frac{g_{A}}{2} N^{\dagger} \tau_{a} \sigma_{i} N 
-\frac{1}{2} L_{1, A} \left(N^{\dagger} \sigma_{i} N\right)
\left(N^{\dagger} \tau_{a} N\right),
\label{eq:axial_current}
\\[10pt]
A_{i,0}={}& -\frac{g_{A,0}}{2} N^{\dagger}  \sigma_{i} N 
 + L_{2, A}\left(N^{\dagger} \sigma_{i} N\right)
\left(N^\dagger  N\right),
\label{eq:isoscalar_axial_current_2}
\\[10pt]
S_0 ={}& g_{S,0} N^{\dagger}  N 
-\frac{1}{2} \widetilde{C}_0 \left(N^{\dagger} N\right)\left(N^{\dagger}  N\right),
\nonumber \\
&
-\frac{1}{2}\widetilde{C}_1\left(N^{\dagger} \sigma_i N\right)\left(N^{\dagger} \sigma_i N\right),
\label{eq:scalar_isoscalar_current_2}
\\[10pt]
S_a ={}& g_{S,3} N^{\dagger} \tau_a  N 
- \frac{1}{2}\widetilde{C}_V \left(N^{\dagger} \tau_a N\right)\left(N^{\dagger}  N\right),
\label{eq:scalar_isovector_current_2}
\\[10pt]
T_{ij,0} ={}& \frac{g_{T,0}}{2} \epsilon_{ijk} N^{\dagger} \sigma_{k} N  
-\frac{1}{2} L_{2, T} \epsilon_{ijk} \left(N^{\dagger} \sigma_k N\right)\left(N^{\dagger} N\right),
\label{eq:tensor_isoscalar_current}
\\[10pt]
T_{ij;a} ={}& \frac{g_{T,3}}{2} \epsilon_{ijk} N^{\dagger} \tau_{a} \sigma_{k} N \nonumber \\
&-\frac{1}{2} L_{1, T}  \epsilon_{ijk} \left(N^{\dagger} \sigma_k N\right)\left(N^{\dagger} \tau_a N\right).
\label{eq:tensor_isovector_current_2}
\end{align}
Note that each of these terms is Hermitian so no Hermitian conjugation is implied. For the scalar currents, 
$\widetilde{C}_T=\widetilde{C}_{0}-3 \widetilde{C}_{1}$ and $\widetilde{C}_S=\widetilde{C}_{0}+\widetilde{C}_{1}$ in Eq.~\eqref{eq:scalar_isoscalar_current}, and 
$\{\widetilde{C}_0,\widetilde{C}_1\}=\{\overline{C}_S,4\overline{C}_T\}$ of Ref \cite{Krebs:2020plh}.

\begin{figure*}[!t]
    \centering
    \subfigure[]{
    \includegraphics[width=0.45\textwidth]{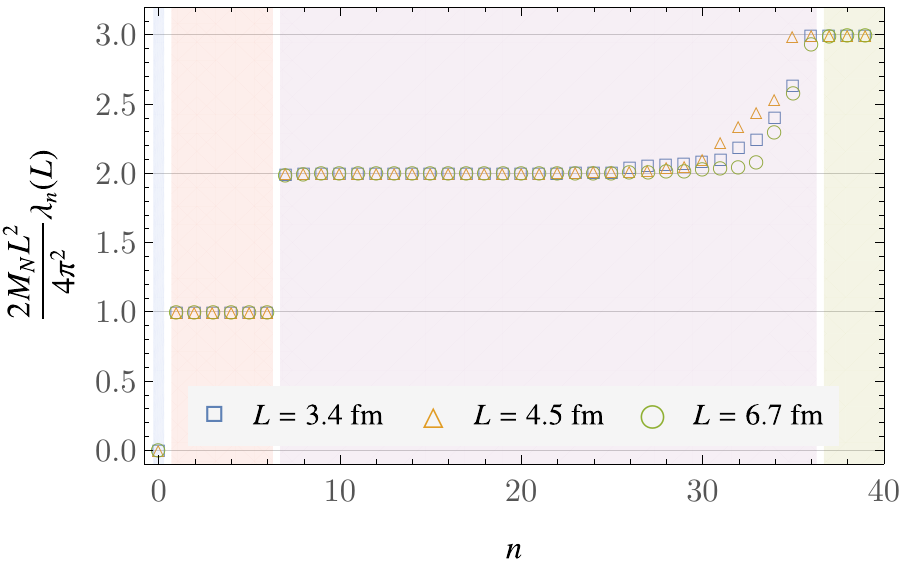}
    \label{fig:evals}}
    \centering
    \subfigure[]{
    \includegraphics[width=0.46\textwidth]{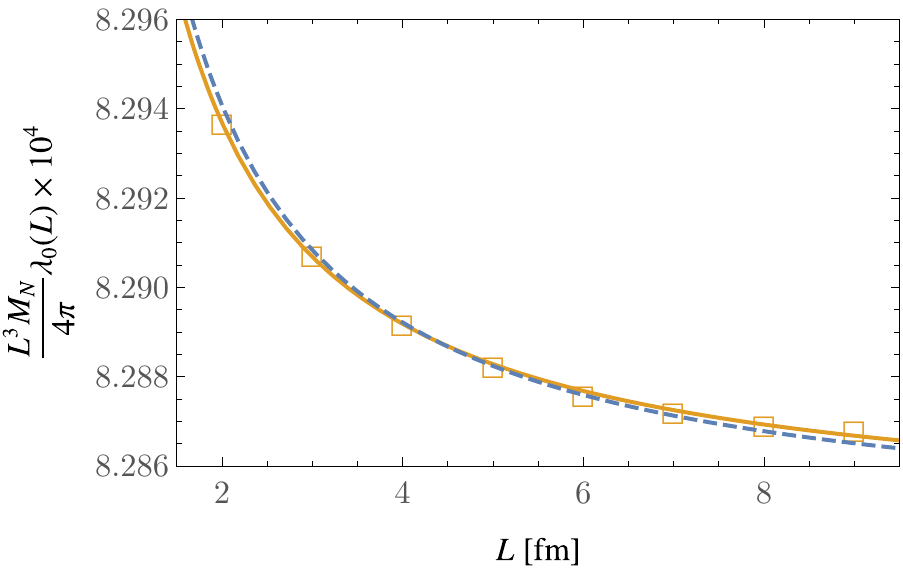}
    \label{fig:luscher}}
    \caption{(a) Energy eigenvalues, $\lambda_n$ for the free two-nucleon system obtained using the SVM in three different volumes, plotted in units of $2M_NL^2/4\pi^2$. The coloured regions indicate the expected multiplicity of eigenvalues. (b) Energy eigenvalues, $\lambda_0$ for a weakly repulsive interaction obtained using the SVM in multiple  volumes. The solid and dashed curves correspond to fits using Eq.~\eqref{eq:luscherexpansion} with either $\{a,c_1\}$ or $\{a\}$ as fit parameters, respectively.}
\end{figure*}

Similarly, the two body part of Eq.~\eqref{eq:Lmagfield} can be written as
\begin{align}
    {\cal L}_{2,\text{EM}} ={}&  -\frac{e}{2} L_{1}\left(N^{\dagger} \sigma\cdot {\bf B} N\right)\left(N^{\dagger} \tau_{3} N\right) 
    \nonumber \\&
    + \frac{e}{2} L_{2} \left(N^{\dagger} \sigma\cdot {\bf B} N\right)\left(N^{\dagger} N\right).
        \label{eq:L2magfield2}
\end{align}

\section{Scattering states for free particles and weak interactions}
\label{app:scattering}

The correlated shifted Gaussian basis is able to accurately describe low-energy finite-volume scattering states as well as localised bound states. To demonstrate this, SVM approximations for non-interacting two particle states are studied in this appendix. 

Figure~\ref{fig:evals} shows the energy eigenvalues obtained for a free two-nucleon system using wavefunctions approximated using the shifted correlated Gaussian basis functions. The eigenvalues are shown in units of $2M_NL^2/4\pi^2$, in which case the expectation is an integer-spaced spectrum. States are symmetric under particle interchange and should be ordered in terms of the sum of the squared momenta, ${\cal N}=|{\bf p}_1|^2+|{\bf p}_2|^2|{\bf p}_1|^2$, and should have degeneracies $1,6,30,\ldots$ corresponding to $\{p_1=(0,0,0), p_2=(0,0,0)\}$ for ${\cal N}=0$, $\{p_1=(1,0,0), p_2=(0,0,0)\}$ and permutations and sign changes for ${\cal N}=1$, 
$\{p_1=(1,1,0), p_2=(0,0,0)\}$ or $\{p_1=(1,0,0), p_2=(0,1,0)\}$ and permutations and sign changes for ${\cal N}=2$, and so on. 
As can be seen from the figure, the SVM is able to cleanly reproduce the low-energy part of the spectrum, including its degeneracies; with further numerical effort this can be extended further. Similarly the free three-particle low-energy spectrum is well reproduced.

The large $L$ asymptotic behaviour of the ground state is given by an expansion of the two-particle quantisation condition derived by L\"uscher \cite{Luscher:1986pf}, namely
\begin{align}
\label{eq:luscherexpansion}
\lambda_{0}=&\frac{4 \pi a}{M_N L^{3}}\left[1-c_{1} \frac{a}{L}+c_{2}\left(\frac{a}{L}\right)^{2}+\ldots\right]+\mathcal{O}\left(L^{-6}\right),
\end{align}
where $a$ is the scattering length and the  geometric coefficients are $c_{1}=-2.837297$ and $c_{2}=+6.375183$. 
Figure \ref{fig:luscher} shows the two-body ground-state energy extracted from the SVM with a small repulsive coupling (i.e., the energy in the isoscalar channel with $C_S=31$~MeV fm$^3$ and $C_T=0$ in Eq.~\eqref{eq:Lagrangian2body}) as a function of volume. Also shown are fits to this data using Eq.~\eqref{eq:luscherexpansion} with the scattering length as a free parameter, as well as fits treating both the scattering length and the geometric constant $c_1$ as free parameters. The latter fit returns a value of $c^{\rm fit}_1=-2.63(19)$ that is consistent with the actual value, showing that the SVM is correctly able to reproduce the expected finite-volume asymptotic behaviour. Similar agreement with the expect asymptotic behaviour is found for small attractive interactions.

\begin{widetext}
\section{Matrix element formulae}
\label{app:MEintegrals}

\def\bcut{\tilde{b}}
\def\qcut{\tilde{q}}

In this section, explicit formulae for the wavefunction integrals in Eqs.~\eqref{eq:Nint} and ~\eqref{eq:Hint} are provided. In the expressions below, $\Psi_L^\text{sym}(A_i,B_i,{\bf d}_i;{\bf{x}})$ is a symmetric $n$-body Gaussian wavefunction term defined in Eq.~\eqref{eq:symwf}. 
The normalisation integral of Eq.~\eqref{eq:Nint}, corresponding to a cross-term between such $n$-body Gaussian wavefunction terms labelled by the subscripts $i$ and $j$ respectively, can be expressed as
\begin{align}
    [\mathbb{N}]_{ij} \equiv &\int \Psi_L^\text{sym}(A_i,B_i,{\bf d}_i;{\bf{x}})^* \Psi_L^\text{sym}(A_j,B_j,{\bf d}_j;{\bf{x}}) d{\bf{x}}\nonumber\\
    = &\sum_{\mathcal{P},\mathcal{P}'} \prod_{\alpha\in\{x,y,z\}}
    \sqrt{\frac{(2\pi)^{n}}{\text{Det}[C^{(\alpha)}_{i\mathcal{P};j\mathcal{P}'}]}}
    \sum_{{\bf b}^{(\alpha)}}^{|{\bf b}^{(\alpha)}|\le \bcut}
   \text{exp}\Bigg[
    -\frac{1}{2} \Omega^{(\alpha)}_{i\mathcal{P};j\mathcal{P}'}
    \Bigg],\label{eq:N}
\end{align}
where 
\begin{equation}
    \mathbf{\Xi}^{(\alpha)}_{i\mathcal{P};j\mathcal{P}'}
    = LA^{(\alpha)}_{i\mathcal{P}}\cdot {\bf b}^{(\alpha)}+B^{(\alpha)}_{i\mathcal{P}}\cdot(L{\bf b}^{(\alpha)}+{\bf d}^{(\alpha)}_{i\mathcal{P}})+B_{j\mathcal{P}'}\cdot{\bf d}^{(\alpha)}_{j\mathcal{P}'}\,,
\end{equation}
\begin{align}
\Omega^{(\alpha)}_{i\mathcal{P};j\mathcal{P}'}
={}& (L{\bf b}^{(\alpha)})\cdot A^{(\alpha)}_{i\mathcal{P}}\cdot(L{\bf b}^{(\alpha)})
    +(L{\bf b}^{(\alpha)}+{\bf d}^{(\alpha)}_{i\mathcal{P}})\cdot B^{(\alpha)}_{i\mathcal{P}} \cdot (L{\bf b}^{(\alpha)}+{\bf d}^{(\alpha)}_{i\mathcal{P}}) 
    +{\bf d}^{(\alpha)}_{j\mathcal{P}'}\cdot B^{(\alpha)}_{j\mathcal{P}'} \cdot {\bf d}^{(\alpha)}_{j\mathcal{P}'}
    -  \mathbf{\Xi}^{(\alpha)}
    \cdot 
    [C^{(\alpha)}_{i\mathcal{P};j\mathcal{P}'}]^{-1} 
    \cdot
    \mathbf{\Xi}^{(\alpha)}\,,
\end{align}
and
\begin{equation}
    C^{(\alpha)}_{i\mathcal{P};j\mathcal{P}'} = A_{i\mathcal{P}}^{(\alpha)}+A_{j\mathcal{P}'}^{(\alpha)}+B_{i\mathcal{P}}^{(\alpha)}+B_{j\mathcal{P}'}^{(\alpha)},
\end{equation}
and where superscripts $^{(\alpha)}$ and subscripts $_{\mathcal{P}}$ on the wavefunction parameters $A$, $B$, and ${\bf d}$ extract the components of the parameters corresponding to the $\alpha$ direction, and permute the parameters for each of the $n$ particles by the permutation $\mathcal{P}$, respectively. 
In the above equations, ${\bf b}^{(\alpha)}$ is a length-$n$ vector; summing over all $n$-vectors corresponds to enforcing periodicity. In practice the infinite sum is truncated to vectors with a maximum norm $\bcut$, and in the numerical calculations in this work the cut is initially taken to be $\bcut =3$ for each term, and is iteratively increased until the fractional change in the total summed expression from adding an additional term is less than $10^{-10}$.

The spatial integrals involved in the Hamiltonian matrix elements are separated into the kinetic and two and three-body potential terms as
\begin{equation}
    \mathbb{H}=\langle \chi_h | \mathbb{K}+(C_0 + C_1 \sigma\cdot\sigma)\mathbb{V}_2+D_0\mathbb{V}_3 | \chi_h \rangle,
\end{equation}
where the spatial integrals for each of $\mathbb{K}$, $\mathbb{V}_2$, and  $\mathbb{V}_3$ can be performed independently.
The integral for the two-body potential term, again for $n$-body Gaussian wavefunction terms labelled by $i$ and $j$, is
\begin{align}
    [\mathbb{V}_2]_{ij} &\equiv \sum_{a<b}^{n} \int \Psi_L^\text{sym}(A_i,B_i,{\bf d}_i;{\bf{x}})^* g_\Lambda({\bf x}_{a}-{\bf x}_{b},L)\Psi_L^\text{sym}(A_j,B_j,{\bf d}_j;{\bf{x}}) d{\bf{x}} \nonumber \\
    & = \sum_{\mathcal{P},\mathcal{P}'}\sum_{a<b}^{n}\prod_{\alpha\in\{x,y,z\}} \sqrt{\frac{(2\pi)^{n}}{\text{Det}[C^{(\alpha)}_{i\mathcal{P};j\mathcal{P}'}]}}\sqrt{\frac{\widetilde{C}^{(\alpha)}_{i\mathcal{P};j\mathcal{P}'}}{\widetilde{C}^{(\alpha)}_{i\mathcal{P};j\mathcal{P}'}+2\rho}}
    \sum_{{\bf b}^{(\alpha)}}^{|{\bf b}^{(\alpha)}|\le \bcut}
    \text{exp}\Bigg[ 
    -\frac{1}{2} \Omega^{(\alpha)}_{i\mathcal{P};j\mathcal{P}'}\Bigg]
    \nonumber \\
    & \qquad \times \sum_{q^{(\alpha)}=-\qcut}^{\qcut}
    \text{exp}\Bigg[- \frac{\rho\,\widetilde{C}^{(\alpha)}_{i\mathcal{P};j\mathcal{P}'} }{\widetilde{C}^{(\alpha)}_{i\mathcal{P};j\mathcal{P}'}+2\rho}\left([(C^{(\alpha)}_{i\mathcal{P};j\mathcal{P}'})^{-1} \cdot \mathbf{\Xi}^{(\alpha)}]_a - [(C^{(\alpha)}_{i\mathcal{P};j\mathcal{P}'})^{-1} \cdot \mathbf{\Xi}^{(\alpha)}]_b - Lq^{(\alpha)}\right)^2
    \Bigg],\label{eq:v2}
\end{align}
where the constant $\rho = \frac{1}{2 r_0^2} = \frac{\Lambda^2}{2}$ is a re-scaling of the Gaussian regulator parameter, and
\begin{equation}
    \widetilde{C}^{(\alpha)}_{i\mathcal{P};j\mathcal{P}'} = \left([C^{(\alpha)}_{i\mathcal{P};j\mathcal{P}'}]^{-1}_{aa}+[C^{(\alpha)}_{i\mathcal{P};j\mathcal{P}'}]^{-1}_{bb}-[C^{(\alpha)}_{i\mathcal{P};j\mathcal{P}'}]^{-1}_{ab}-[C^{(\alpha)}_{i\mathcal{P};j\mathcal{P}'}]^{-1}_{ba}\right)^{-1},
\end{equation}
where $[V]_a$ denotes the $a$th component of a vector $V$, and $[M]^{-1}_{ab}$ denotes the $(a,b)$ component of the matrix $M^{-1}$. 
In this expression $q^{(\alpha)}$ is an integer; combined with the sum over ${\bf b}^{(\alpha)}$, summing over all values of $q^{(\alpha)}$ corresponds to enforcing periodicity. 
In practice the infinite sum is truncated to integers with absolute value less than $\qcut=40$; the fractional change in the total summed expression from adding an additional term is less than $10^{-10}$.

The relevant integral for the three-body potential term, for $n$-body Gaussian wavefunction terms labelled by $i$ and $j$, can be expressed as
\begin{align}
    [\mathbb{V}_3]_{ij} &\equiv \sum_{a\ne b\ne c}^{\rm cyc} \int \Psi_L^\text{sym}(A_i,B_i,{\bf d}_i;{\bf{x}})^* g_\Lambda({\bf x}_{a}-{\bf x}_{b},L)g_\Lambda({\bf x}_{b}-{\bf x}_{c},L)\Psi_L^\text{sym}(A_j,B_j,{\bf d}_j;{\bf{x}}) d{\bf{x}} \nonumber \\
     =& \sum_{\mathcal{P},\mathcal{P}'}\sum_{a\ne b\ne c}^{\rm cyc} \prod_{\alpha\in\{x,y,z\}}
    \sqrt{\frac{(2\pi)^{n}}{\text{Det}[{\widehat{C}}^{(\alpha)}_{i\mathcal{P};j\mathcal{P}'}]}} 
\text{exp}\left[-\frac{1}{2}\left(
    {\bf d}^{(\alpha)}_{i\mathcal{P}} \cdot B^{(\alpha)}_{i\mathcal{P}} \cdot {\bf d}^{(\alpha)}_{i\mathcal{P}} 
    + {\bf d}^{(\alpha)}_{j\mathcal{P}'} \cdot B^{(\alpha)}_{j\mathcal{P}'} \cdot {\bf d}^{(\alpha)}_{j\mathcal{P}'}
    \right)\right] 
    \nonumber \\
    & \qquad
\times \sum_{{\bf b}^{(\alpha)}}^{|{\bf b}^{(\alpha)}|\le \bcut}
\text{exp}\left[
-\frac{1}{2} \left(
(L{\bf b}^{(\alpha)})\cdot (A^{(\alpha)}_{i\mathcal{P}}+B^{(\alpha)}_{i\mathcal{P}}) \cdot (L {\bf b}^{(\alpha)})
+   2 {\bf d}^{(\alpha)}_{i\mathcal{P}} \cdot B^{(\alpha)}_{i\mathcal{P}} \cdot (L {\bf b}^{(\alpha)}) 
     -\mathbf{\Xi}^{(\alpha)} \cdot 
     [{\widehat{C}}^{(\alpha)}_{i\mathcal{P};j\mathcal{P}'}]^{-1}
     \cdot \mathbf{\Xi}^{(\alpha)}
     \right)\right]
     \nonumber \\
     & \qquad
     \times
     \sum_{q^{(\alpha)}=-\qcut}^{\qcut}
    \text{exp}\Bigg[
    -\frac{L^2}{r_0^2} q^{(\alpha)2}
    +\frac{q^{(\alpha)2}L^2}{2r_0^4}
    \mathfrak{P}_v^{[a, b]} \cdot 
    [{\widehat{C}}^{(\alpha)}_{i\mathcal{P};j\mathcal{P}'}]^{-1}
    \cdot \mathfrak{P}_v^{[a, b]}
    + \frac{q^{(\alpha)}L}{r_0^2} \mathbf{\Xi}^{(\alpha)} \cdot 
    [{\widehat{C}}^{(\alpha)}_{i\mathcal{P};j\mathcal{P}'}]^{-1} 
    \cdot \mathfrak{P}_v^{[a, b]}
    \Bigg]
    \nonumber\\
    & \qquad
    \times
     \sum_{t^{(\alpha)}=-\qcut}^{\qcut}
    \text{exp}\Bigg[
    -\frac{L^2}{r_0^2} t^{(\alpha)2}
    +\frac{t^{(\alpha)2}L^2}{2r_0^4}*
    \mathfrak{P}_v^{[b, c]} \cdot 
    [{\widehat{C}}^{(\alpha)}_{i\mathcal{P};j\mathcal{P}'}]^{-1}
    \cdot \mathfrak{P}_v^{[b, c]}
    + \frac{t^{(\alpha)}L}{r_0^2} \mathbf{\Xi}^{(\alpha)} \cdot 
    [{\widehat{C}}^{(\alpha)}_{i\mathcal{P};j\mathcal{P}'}]^{-1} 
    \cdot \mathfrak{P}_v^{[b, c]}
    \Bigg] 
    \nonumber\\& \hspace{22mm}
    \times \text{exp}\left[\frac{t^{(\alpha)}q^{(\alpha)}L^2}{r_0^4}\mathfrak{P}_v^{[b, c]}\cdot [{\widehat{C}}^{(\alpha)}_{i\mathcal{P};j\mathcal{P}'}]^{-1}
    \cdot\mathfrak{P}_v^{[a, b]})\right]
    \,,\label{eq:v3}
\end{align}
where $\sum_{a\ne b\ne c}^{\rm cyc}$ indicates a sum over cyclic permutations of  particles $a$, $b$ and $c$, 
\begin{align}
    {\widehat{C}}^{(\alpha)}_{i\mathcal{P};j\mathcal{P}'} =
    C^{(\alpha)}_{i\mathcal{P};j\mathcal{P}'} + \frac{1}{r_0^2} \left(\mathfrak{P}_m^{[a, b]} + \mathfrak{P}_m^{[b, c]} \right),
\end{align}
and vector and matrix projectors are defined component-wise as
\begin{align}
    [\mathfrak{P}_v^{[a, b]}]_c &= \delta_{ac} - \delta_{bc} \,,\\
    [\mathfrak{P}_m^{[a, b]}]_{cd} &=
    \delta_{cd}(\delta_{ac} +\delta_{bc}) -
    \delta_{ac}\delta_{bd}-\delta_{ad}\delta_{bc}\,.
\end{align}
As in Eqs.~\eqref{eq:N} and \eqref{eq:v2}, the sums over ${\bf b}^{(\alpha)}$, $q^{(\alpha)}$, and $t^{(\alpha)}$ together enforce periodicity. In numerical evaluations of Eq.~\eqref{eq:v3}, the same cut procedure for fixing $\bcut$ and $\qcut$ are used as for the evaluations of the previous expressions.

Finally, the integral for the kinetic operator, for $n$-body Gaussian wavefunction terms labelled by $i$ and $j$, is
\begin{align}
    [\mathbb{K}]_{ij} &\equiv -\frac{1}{2M_N} \sum_{a=1}^n \int \Psi_L^\text{sym}(A_i,B_i,{\bf d}_i;{\bf{x}})^*  \nabla_a^2\Psi_L^\text{sym}(A_j,B_j,{\bf d}_j;{\bf{x}}) d{\bf{x}} \nonumber \\
    & = 
   \frac{1}{2M_N}  \sum_{\mathcal{P},\mathcal{P}'}
     \sum_{\alpha\in\{x,y,z\}} \sqrt{\frac{(2\pi)^{n}}{\text{Det}[C^{(\alpha)}_{i\mathcal{P};j\mathcal{P}'}]}}
      \sum_{{\bf b}^{(\alpha)}}^{|{\bf b}^{(\alpha)}|\le \bcut}
      \Theta^{(\alpha)}_{i\mathcal{P};j\mathcal{P}'}
      \text{exp}\Bigg[
    -\frac{1}{2} \Omega^{(\alpha)}_{i\mathcal{P};j\mathcal{P}'}
    \Bigg]
   \nonumber  \\ & \hspace{6cm}
    \times\prod_{\beta\in\{x,y,z\}}^{\beta\ne\alpha}
    \sqrt{\frac{(2\pi)^{n}}{\text{Det}[C^{(\beta)}_{i\mathcal{P};j\mathcal{P}'}]}}
      \sum_{{\bf b}^{(\beta)}}^{|{\bf b}^{(\beta)}|\le \bcut}
      \text{exp}\Bigg[
    -\frac{1}{2} \Omega^{(\beta)}_{i\mathcal{P};j\mathcal{P}'}\Bigg],
\end{align}
where $\hbar=1$ is used and 
\begin{align}
     \Theta^{(\alpha)}_{i\mathcal{P};j\mathcal{P}'} =&
      {\rm Tr}\left[[(A_{i{\cal P}}^{(\alpha)} + B_{i{\cal P}}^{(\alpha)})
      \cdot \left[C^{(\alpha)}_{i\mathcal{P};j\mathcal{P}'}\right]^{-1}
      \cdot (A_{j{\cal P}'}^{(\alpha)} + B_{j{\cal P}'}^{(\alpha)})\right]
      \nonumber\\
      &-\Bigg|
      (A_{j{\cal P}'}^{(\alpha)} + B_{j{\cal P}'}^{(\alpha)})
      \cdot
       \left[C^{(\alpha)}_{i\mathcal{P};j\mathcal{P}'}\right]^{-1}
       \cdot
       \left(
        B_{i{\cal P}}^{(\alpha)}\cdot(L{\bf b}^{(\alpha)}+ {\bf d}_{i{\cal P}}^{(\alpha)}) + A_{i{\cal P}}^{(\alpha)}\cdot(L{\bf b}^{(\alpha)})
       \right)
       \nonumber\\ & \hspace{5cm}
       - (A_{i{\cal P}}^{(\alpha)} + B_{i{\cal P}}^{(\alpha)})
       \cdot 
       \left[C^{(\alpha)}_{i\mathcal{P};j\mathcal{P}'}\right]^{-1}
       \cdot B_{j{\cal P}'}^{(\alpha)} \cdot {\bf d}_{j{\cal P}'}^{(\alpha)})
      \Bigg|^2.
\end{align}

\end{widetext}

\bibliography{bib.bib}
\end{document}